\documentclass[10pt,conference,letterpaper]{IEEEtran}
\usepackage{graphicx}
\usepackage{booktabs}
\usepackage{balance}  
\usepackage{multirow}
\usepackage{ifpdf}
\usepackage{times,amsmath,epsfig}
\usepackage{epsfig}
\usepackage{amssymb}
\usepackage{epstopdf}
\usepackage{hyperref}
\usepackage{subfigure}
\usepackage{caption}
\usepackage{color}
\usepackage{url}
\usepackage[noend,linesnumbered,ruled,vlined]{algorithm2e}
\usepackage{mathrsfs}

\newtheorem{definition}{Definition}
\newtheorem{theorem}{Theorem}
\newtheorem{lemma}{Lemma}

\newcommand{\stitle}[1]{\vspace{.5ex}\noindent{\bf #1}}

\newcommand{\wrt}{\emph{w.r.t.}\xspace}

\newcommand{\eop}{\hspace*{\fill}\mbox{$\Box$}}
\newcounter{example}
\newenvironment{example}{\refstepcounter{example}{\vspace{1ex} \noindent\bf Example \arabic{example}:} }{\eop \vspace{1ex}}

\newcommand{\brm}{\textsc{surge}\xspace}
\newcommand{\bpm}{\textsc{cSpot}\xspace}
\newcommand{\mer}{\textsc{mer}\xspace}
\newcommand{\eat}[1]{}
\definecolor{orange}{RGB}{0,0,255}

\begin{document}

\title{SURGE: Continuous Detection of Bursty Regions Over\\ a Stream of Spatial Objects}

\author{{Kaiyu Feng$^{1, 2}$, Tao Guo{$^{2}$}, Gao Cong{$^{2}$}, Sourav~S.~Bhowmick$^{2}$, Shuai Ma{$^{3}$}}\\
$^1$ LILY, Interdisciplinary Graduate School. Nanyang Technological University, Singapore\\
$^2$ School of Computer Science and Engineering, Nanyang Technological University, Singapore \\
$^3$ SKLSDE, Beihang University, China \\
\{kfeng002@e., tguo001@e., gaocong@, assourav@\}ntu.edu.sg, mashuai@buaa.edu.cn
}





\maketitle

\begin{abstract}
With the proliferation of mobile devices and location-based services, continuous generation of massive volume of streaming spatial objects (i.e., geo-tagged data) opens up new opportunities to address real-world problems by analyzing them. In this paper, we present a novel \emph{continuous bursty region detection} (\brm) problem that aims to continuously detect a \textit{bursty region} of a given size in a specified geographical area from a stream of spatial objects. Specifically, a bursty region shows maximum spike in the number of spatial objects in a given time window. The \brm problem is useful in addressing several real-world challenges such as surge pricing problem in online transportation and disease outbreak detection. To solve the problem, we propose an exact solution and two approximate solutions, and the approximation ratio is $\frac{1-\alpha}{4}$ in terms of the burst score, where $\alpha$ is a parameter to control the burst score. We further extend these solutions to support detection of t\textit{op-$k$ bursty regions}. Extensive experiments with real-world data are conducted to demonstrate the efficiency and effectiveness of our solutions.

\end{abstract}

\vspace{-1ex} \section{Introduction} \label{sec:intro}
People often share geo-tagged messages through many social services like \textit{Twitter} and \textit{Facebook}. Each geo-tagged data is associated with a timestamp, a geo-location, and a set of attributes (e.g., tweet content). In this paper, we refer to them as \textit{spatial objects}. With the proliferation of GPS-enabled mobile devices and location-based services, the amount of such spatial objects (e.g., geo-tagged tweets and trip requests using \textit{Uber}) is growing at an explosive rate. Their real-time nature coupled with multi-faceted information and rapid arrival rate in a streaming manner open up new opportunities to address real-world problems. For example, consider the following problems.

\begin{example}\label{example:zika} The world regularly faces the challenge of tackling a variety of virus epidemics such as \textsc{sars}, \textsc{mers}, Dengue, and Ebola. Most recently, the outbreak of mosquito-borne Zika virus started in Brazil in 2015. 
Hence, the Center for Disease Control and Prevention needs to continuously monitor different areas for possible Zika outbreak and issue alerts to people who are traveling to or living  in regions affected by Zika. Since early detection of such outbreak is paramount, how can we identify potential Zika-affected region(s) in real time?

One strategy to address this issue is to continuously monitor geo-tagged tweets (i.e., spatial objects) coming out of a specific area (e.g., Florida) and detect regions where there are sudden bursts in tweets related to Zika (e.g., containing Zika-related keywords) in real time. Observe that these ``bursty regions'' are dynamic in nature. However, it is computationally challenging to continuously monitor massive streams of spatial objects and detect bursty regions in real time.
\end{example}

\begin{example}\label{example:uber} Online transportation network companies such as \textit{Uber}, \textit{Lyft}, and \textit{Didi Dache} have disrupted the traditional transportation model and have gained tremendous popularity among consumers\footnote{\scriptsize In 2017, Uber is available in over 81 countries and 570 cities worldwide.}. Consumers can submit a trip request through their mobile apps. If a nearby driver accepts the request, he will pickup the consumer.

\begin{figure}[t]
  \centering
  \includegraphics[width=\linewidth, height=3cm]{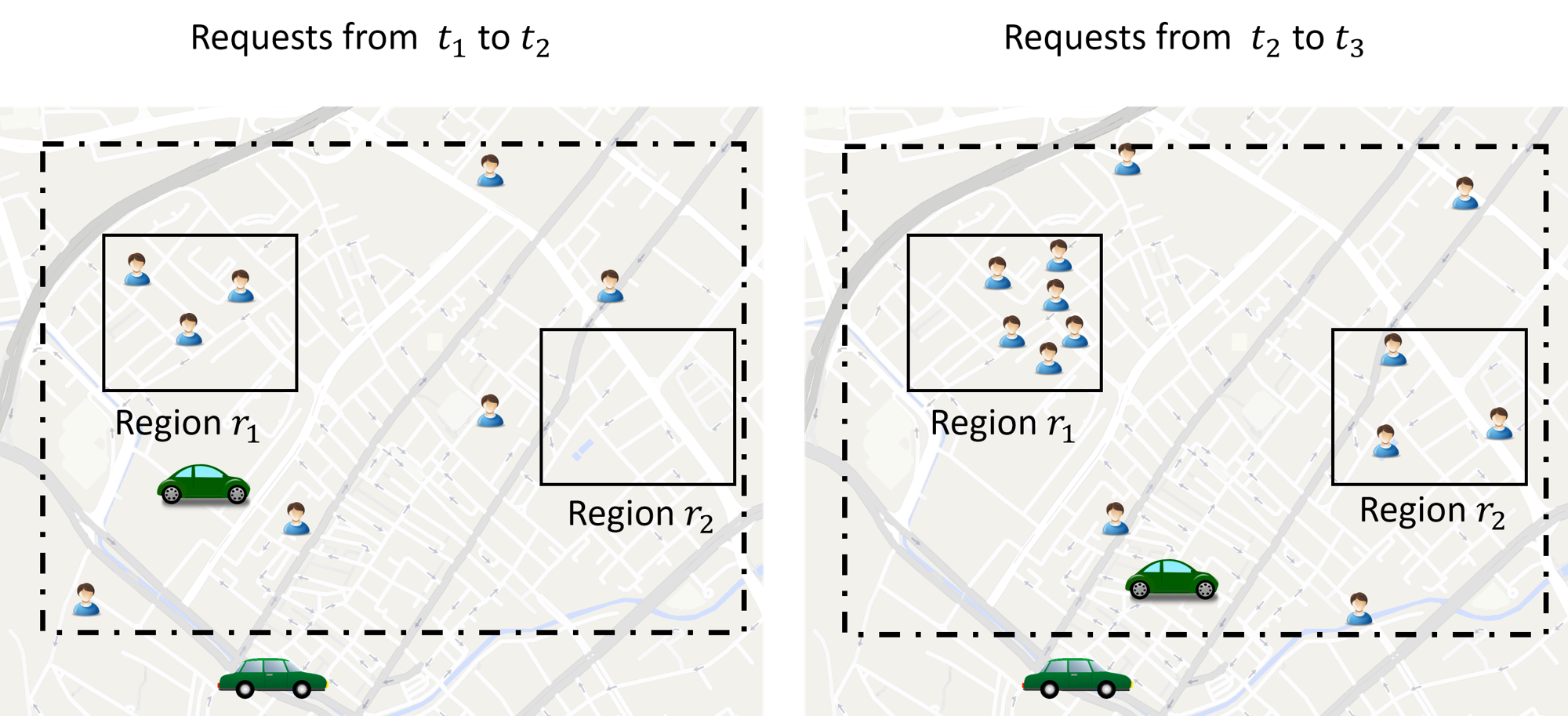}
 \caption{Motivating example.}\label{fig:uberexample}
\vspace{-4ex} \end{figure}

Although this disruptive model has benefited many drivers and consumers, the latter may have to wait for a long time for a car when the number of car requests significantly surpasses the supply of nearby drivers. Clearly, it is beneficial to both passengers and drivers if we can notify idle drivers in real time whenever there is a sudden burst in consumer demand in areas of interest to them.  An additional benefit to the drivers is that the trip fare may be increased due to the ``surge pricing'' policy~\footnote{ \scriptsize For example, the price increased 10X on new year's eve in 2016 in the United States (\url{www.geekwire.com/2016/customers-complain-uber-prices-surge-near-10x-new-years-eve/})} where the  companies may increase a trip price significantly when demand is high. For instance, consider Figure~\ref{fig:uberexample}, which shows the trip requests in two time windows $[t_1, t_2]$ and $[t_2, t_3]$. Suppose a driver is only interested in the area shown by dashed rectangle to pick up passengers. Observe that there is a burst of trip requests in regions $r_1$ and $r_2$ (both increased by 3). If the app can notify the driver in real time about these two regions, then he can move in there to pickup potential passengers. Note that such soaring demand is not always predictable as it may not only occur during holidays or periodic events (e.g., new year's eve) but also due to unpredictable events such as subway disruption, concerts, road accident, inclement weather, and terrorist attack.
\end{example}

There are two common themes in the two examples. First, we need to continuously monitor a large volume of spatial objects (e.g., trip requests and geo-tagged tweets) to detect in real time one or more regions that show relatively large spike in the number of spatial objects (i.e., bursty region) in a given time window. Second, a user needs to specify as input the size $a\times b$ of rectangular-shaped bursty region that one wishes to detect. For instance, in Example~\ref{example:uber} different drivers may prefer bursty regions of different sizes according to their convenience. 

In this paper, we refer to the problem embodied in the aforementioned motivating examples as \emph{continuou\textbf{s} b\textbf{u}rsty \textbf{r}e\textbf{g}ion d\textbf{e}tection} (\brm) problem. Specifically, given a  region size $a\times b$ and an area $A$, the aim of the \brm problem is to continuously detect a region of the specified size in $A$ that demonstrates the \textit{maximum burstiness} from a stream of spatial objects. To model the burstiness of a region, we propose a general function based on the sliding window model. We also extend our \brm problem to detect \textit{top}-$k$ bursty regions as in certain applications one may be interested in a list of such regions. 

The \brm problem and its \textit{top-$k$} variant are challenging as we need to handle rapidly arriving spatial objects in high volumes to efficiently detect and maintain bursty regions. For example, 10 million geo-tagged tweets are generated each day in \textit{Twitter}\footnote{\scriptsize \url{https://www.mapbox.com/blog/twitter-map-every-tweet/}}. 
As we shall see later, it is prohibitively expensive to recompute bursty regions frequently. 

In this paper, we first propose an exact solution called \textit{cell-}\textsc{cSpot} to keep track of the bursty region over sliding windows. Specifically, we first reduce the \brm problem to \emph{\textbf{c}ontinuou\textbf{s} \textbf{b}ursty p\textbf{o}int de\textbf{t}ection (\bpm)} problem. Then we propose a cell-based algorithm to continuously detect the bursty point. 
It takes $O(|c_{max}|^2 + \log n)$ time to process a new arriving spatial object on average, where $|c_{max}|$ is the maximum number of objects that we search inside a cell, and $n$ is the number of indexed rectangle objects.

Although \textit{cell-}\textsc{cSpot }can address the \brm problem efficiently in several scenarios, it becomes inefficient as $|c_{max}|$ increases (e.g., the sliding windows get larger, the region size gets larger, or the arrival rate of the spatial objects increases). To address this we further propose two approximate solutions, namely \textsc{gap-surge} and \textsc{mgap-surge}, with an $O(\log n)$ time complexity to process a spatial object. The approximation ratio is bounded by $\frac{1-\alpha}{4}$, where $\alpha \in [0,1)$ is a parameter used in the \textit{burst score function}. Last, we show that our proposed solutions can be elegantly extended to continuously detect top-$k$ bursty regions. Our experiments reveal that our proposed solutions can handle streams with up to 10 millions spatial objects arrived per day. 

In summary, this paper makes the following contributions:

\vspace{.5ex}
(1) We propose a novel \emph{continuou\textbf{s} b\textbf{u}rsty \textbf{r}e\textbf{g}ion d\textbf{e}tection} (\brm) problem for continuously detecting bursty regions in a specified area from a stream of spatial objects. (Section \ref{sec:definition})

\vspace{.5ex}
(2)  We present an exact solution (\textit{cell-}\textsc{cSpot}) and two approximate solutions (\textsc{gap-surge} and \textsc{mgap-surge}) to address the \brm problem (Sections~\ref{sec:excat} and ~\ref{sec:approximate}). We further extend these solutions to keep track of top-$k$ bursty regions efficiently (Section~\ref{sec:topk}).

\vspace{.5ex}
(3)  We conduct experiments with real-world datasets to show the efficiency of our proposed solutions. All solutions are efficient in real time. Moreover, \textsc{gap-surge} and \textsc{mgap-surge}  scale well w.r.t. high arrival rate  while the returned regions have competitive burst scores. The extended versions can also detect top-$k$ bursty regions efficiently in real time. (Section~\ref{sec:exp}).

The proofs of lemmas and theorems are given in Appendix~\ref{appendix:proofs}.

\vspace{-1ex} \section{Related Work}\label{sec:related}

\stitle{Burst detection}. Our \brm problem is related to the problem of
detecting bursty patterns and topics. A host of work has been done to detect
temporal bursts \cite{kleinberg2003bursty, zhu2003efficient, fung2005parameter,bulut2005unified, wang2007mining}. A collection of proposals focus on detecting bursty features (represented by probability distribution of words)\cite{kleinberg2003bursty, fung2005parameter, wang2007mining}. The other work focuses on detecting a timespan over the stream such that its aggregate is larger than a threshold \cite{zhu2003efficient, bulut2005unified}. All these burst detection problems are different from our \brm problem as they disregard the spatial information when detecting the temporal bursts.

Most germane to our work are efforts on exploring spatial-temporal bursts~\cite{mathioudakis2010identifying, lappas2012spatiotemporal, zhang2016geoburst} albeit from different aspects. Mathioudakis et al. \cite{mathioudakis2010identifying} study the problem of identifying notable spatial burst out of a collection of user generated information. They divide the space into cells, and recognize two states for each cell, namely ``bursty'' and ``non-bursty''. 
additive cost function.
Our \brm problem differs from it in two key aspects. First, the spatial burst is identified as a cell in the grid whereas the bursty region in \brm can be located at any position. Second, the solution developed in~\cite{mathioudakis2010identifying} is designed for data stored in a data warehouse, and it cannot be deployed or adapted to solve the \brm problem. 
Lappas et
al. \cite{lappas2012spatiotemporal} study the problem of identifying a combination of a temporal interval and a geographical region with unusual high frequency for a term from a set of geo-tagged text streams.
Its problem setting is different from ours: Lappas et al. 
\cite{lappas2012spatiotemporal} takes as input a set of text streams with fixed geographical locations, while in our \brm problem, spatial objects arrive as
a stream and an object can be located in any location of the given space. In addition,
the proposed solution can only handle a small number of text streams (tens to
hundreds) due to its high computational complexity. 
Given a geo-tagged tweet stream, Zhang et al. \cite{zhang2016geoburst} aim to continuously detect real-time local event. Specifically, a local event is defined as a cluster of tweets that are semantically coherent and geographically close. For each keyword in a tweet, its burstiness is a linear combination of its temporal burstiness and its spatial burstiness with a balance parammeter $\eta$. The spatiotemporal burstiness of a cluster of tweets is the aggregation of the burstiness of all the keywords in the cluster.
Our problem differs from it in the following aspects. First, the bursty event is identified as a cluster of geo-tagged tweets, while our \brm problem aims to detecting a spatial region. Second, the proposed framework is built over geo-textual stream. The textual content serves as an important feature in their system. Our \brm is applicable to any kind of spatial stream.

\stitle{Dense region search}. Our problem is also related to dense region search over moving objects~\cite{jensen2006effective, ni2007pointwise}. Given a set of moving objects, whose positions are modeled as linear functions in Euclidean space, the dense region search problem aims to find all dense regions at query time $t$. Jensen et al. \cite{jensen2006effective} constraint dense regions to be non-overlapping square-shaped regions of given size, whose density is larger than a user-specified threshold. 
Ni et al. \cite{ni2007pointwise} propose a new definition of dense regions, which may have arbitrary shape and size. 
In the dense region search problem, the positions of the moving objects are modeled as linear functions. Thus the position of each moving object can be computed at any time. 
In contrast, in the \brm problem, the number of the newly-arriving spatial objects and their positions are unknown a priori. Moreover, the density function is different from our burst score function, requiring different techniques to compute the burst score of a given region.

\stitle{Region search}. Our problem is also related to the region search problem. A class of studies aims to find a region of a given size such that the \textit{aggregation score} of the region is maximized\cite{nandy1995unified, choi2012scalable, tao2013approximate,feng2016towards}. Given a set of spatial objects, the \emph{max-enclosing rectangle} (\mer) problem \cite{nandy1995unified} aims to find the position of a rectangle of a given size $a\times b$ such that the rectangle encloses the maximum number of spatial objects.
This problem is systematically investigated as the \emph{maximizing range sum} (\emph{MaxRS}) problem \cite{choi2012scalable, tao2013approximate}. 
Feng et al.\cite{feng2016towards} further study a generalized problem of the \emph{MaxRS} problem, in which the aggregate score function is defined by submodular monotone functions, which include sum. Liu et al. \cite{liu2011subject} study the problem of finding subject oriented top-$k$ hot regions, which can be considerd as a top-k version of the \emph{MaxRS} problem. Cao et al.\cite{cao2014retrieving} study the problem of finding a subgraph of a given size with the maximum aggregation score from a road network. All these aforementioned region search problems focus on static data. Moreover, the idea of invoking the approach designed for the region search problem whenever a object enters or leaves the sliding windows is prohibitively expensive (We will elaborate on this in Section~\ref{sec:handlestream}).

Our work is closely related to the recent efforts on continuous \emph{MaxRS} problem~\cite{amagata2016monitoring, Hussain2017Towards}. Amagata et al. \cite{amagata2016monitoring} propose the problem of monitoring the \emph{MaxRS} region over spatial data streams. Specifically, given a stream of weighted spatial objects, the continuous \emph{MaxRS} problem aims to monitor the location of a rectangle of a size $a\times b$ such that the sum of the weights of the objects covered by the rectangle is maximized. In the proposed algorithm, a grid is imposed over the space, whose granularity is independent from the size of the query rectangle. For each spatial object in the stream, they generate a rectangle of a size $a\times b$ whose center is located at the spatial object. The generated rectangle is mapped to the cells with which it overlaps. For each cell, they maintain a graph where each node in the graph is a rectangle mapped to this cell, and two nodes are connected by a directed edge if they overlap with each other. The graph is used to handle the updates of the stream. For each rectangle in the cell,  they maintain an upper bound to determine when to invoke the sweep-line algorithm \cite{nandy1995unified} to find the most overlapped region inside the rectangle. With the maintained upper bounds, they use a branch-and-bound algorithm to reduce the search space. 
The difference of the \brm problem from the continuous \emph{MaxRS} problem is that the burst score of the \brm problem is defined over two consecutive sliding windows, and spatial objects in different windows contribute differently to the burst score. Though their solution cannot be directly applied to solve the \brm problem, we can adapt their solution with some modifications for the \brm problem. The details of the modification are reported in Appendix~\ref{appendix:evaluatedmethods}. 
One issue of this solution is that they need to maintain a graph for each cell with a space cost of $O(n^2)$, where $n$ is the number of rectangle objects that are mapped to the cell. When the number of objects mapped to a cell is large, the space cost could be extremely high. We will show in Section~\ref{sec:exactexp} that our proposed solutions outperform the \emph{aG2} algorithm for the \brm problem. Hussain et al. \cite{Hussain2017Towards} investigates the \emph{MaxRS} problem on the trajectories of moving objects. Given the trajectories of a set of moving points, they aim to maintain the result of the \emph{MaxRS} problem at any time instant. Its problem setting is different from ours: it takes as input the trajectories of a set of fixed number moving objects, while in our problem, the number of spatial objects in the sliding windows may vary with time and the positions of the new arrived objects are unknown a priori.

\stitle{Data stream management}. Our work is also related to data stream management. There has been a long stream of work on various aspects of  data streams since the last decade. Some examples are stream clustering~\cite{aggarwal2003framework, o2002streaming}, stream join processing~\cite{das2003approximate}, and stream summarization~\cite{cormode2005improved}.
Since most of these studies focus on general data streams, we only review the work that involves spatial information. Given a stream of spatial-textual objects, \cite{wang2014selectivity} aims to estimate the cardinality of a spatial keyword query on objects seen so far. A host of work has also been done to study content-based publish/subscribe systems~\cite{wang2016skype, chen2015temporal, hu2015location, li2013location, chen2013efficient, wang2015ap} over spatial object streams. In these systems, streaming published items are delivered to the users with matching interests. However, none of these studies consider the problem of detecting bursty regions.

\stitle{Spatial outlier detection}. Lastly, our work is also related to spatial outlier detection \cite{lu2003algorithms, zhao2003detecting, kou2006spatial}. 
Lu et al. \cite{lu2003algorithms, kou2006spatial} investigate the spatial outlier detection problem over point data. Specifically, given a set of weighted spatial points, 
the spatial outlier detection problem aims to identify top $m$ points such that their weight is greatly different from the average weight of its $k$ nearest neighbors. Zhao et al.\cite{zhao2003detecting} further investigate region outliers detection over meteorological data. All these aforementioned problems focus on static data. Moreover, the outliers are selected from the data points in the spatial outlier detection problem. In contrast, the location of the bursty region in our \brm problem can be located at any position in the space. In addition, the spatial outlier detection problems use a totally different function to evaluate how much a data point is different from its neighbors. Due to these differences, their proposed solutions cannot be adapted to address the \brm problem.

\vspace{-1ex}\section{Problem Statement}\label{sec:definition}
We formally define the \emph{Continuou\textbf{S} B\textbf{U}rsty \textbf{R}e\textbf{G}ion D\textbf{E}tection} (\brm) problem. We begin by defining some terminology.
\subsection{Terminology}
A \textit{spatial object} is represented with a triple $o=\langle w, \rho, t_c\rangle$, where $w$ is the weight of $o$, $\rho$ is a location point with latitude and longitude, and $t_c$ is the creation time of object $o$.
In this paper, we consider a stream of spatial objects. For example, geo-tagged tweets in \textit{Twitter} can be viewed as a stream of spatial objects arriving in the order of creation time. The \emph{weight} of a tweet could be the relevance of its textual content to a set of query keywords. The car requests in \textit{Uber} can also be viewed as a stream of spatial objects arriving in the order of calling time. In this case, the weight could be the passenger number or travel fare.

We next introduce two consecutive time-based sliding windows, namely \textit{current} and \textit{past windows}. Given a window size $|W|$, the \textit{current window}, denoted by $W_c$ is a time period of length $|W|$ that stretches back to a time point $t - |W|$ from present time $t$. The \textit{past window}, denoted by  $W_p$ is a time period of length $|W|$ that stretches back to a time point $t - 2 |W|$ from the time point $t - |W|$.

Given a region $r$ and a sliding window $W$, let $O(r, W)$ be the set of spatial objects which is created in $W$ and located in region $r$, i.e., $O(r,W) = \{o|o.\rho\in r \wedge o.t_c \in W\}$. Let $f(r, W)$ be the summation of weights of objects in $O(r, W)$ normalized by $W$'s length, i.e., $f(r, W) = \frac{\sum_{o\in O(r,W)}o.w}{|W|}$, which is the score of a region $r$ w.r.t. the sliding time window $W$. 

Note that in this paper, for the sake of simplicity, we assume the current window and the past window have the same length $|W|$. However, our proposed solution is equally applicable when the two sliding windows have different lengths.

\subsection{Burst Score}

Intuitively, the \emph{burst score} of a region $r$ reflects the variation in the spatial objects in $r$ in recent period. This motivates us to design the burst score based on the current and past windows.


We first discuss the intuition in designing the \textit{burst score} using Example~\ref{example:uber}. In this scenario, \textit{Uber} drivers are interested in regions in which they have a higher chance to pick up a passenger. 
Obviously, a driver is more likely to find a passenger in a region that contains a large number of requests in the current window, which represents the \textit{significance} of the region. On the other hand, if a region suddenly experiences a surge of requests, which represents the \textit{burstiness} of the region, then it is highly likely that existing drivers in that region may not be able to fulfill this sudden increase in demand. Consequently, a driver will have a higher chance to find a passenger there. 

Thus, we consider the following two factors in our burst score: (a) The score of the region w.r.t. the current window, i.e. $f(r, W_c)$, which measures the \textit{significance}, and (b) the increase in the score of the region between the current window and the past window, i.e., $\max(f(r, W_c) - f(r, W_p), 0)$, which measures the \textit{burstiness}. Note that we use the $max$ function to guarantee that the increase in the score between the current and past windows is always non-negative since we are only interested in increase in the score.

We now formally define the burst score as follows.

\begin{definition}\textbf{\textup{Burst Score}.} Given a region $r$, we define its burst score $\mathcal{S}(r)$ as:
\begin{equation}\label{equ:burstscore}
\mathcal{S}(r) = \alpha \max(f(r, W_c) - f(r, W_p), 0) + (1-\alpha) f(r, W_c),
\end{equation}
where 
$\alpha \in [0, 1)$ is a parameter that balances the significance and the burstiness.
\end{definition}

\subsection{Continuous Bursty Region Detection \\ (SURGE)  Problem}\label{sec:definition:problem}
We are now ready to formally define the \brm problem.

\begin{definition}\textbf{\textup{Continuous Bursty Region Detection (\brm) Problem.}}
Consider a stream of spatial objects $\mathcal{O}$. Let  $q = \langle A, a\times b, |W|\rangle$ be a \brm query where $A$ is a preferred
area, $a\times b$ is the size of the query rectangle, and $|W|$ is the length of the current and past windows. Given such a query $q$, the aim of the \textbf{\brm problem} is to continuously detect the position of the region $r$ of size $a\times b$ in $A$ with the maximum burst score. The region $r$ is referred to as the \textbf{bursty region}.
\end{definition}


\vspace{-1ex}\section{An Exact Solution}\label{sec:excat}

The \brm problem is challenging to address due to the following reasons. First, given a snapshot of the stream, we are required to locate the bursty region in the preferred area $A$. Intuitively, this bursty region can be located at any point and it is prohibitively expensive to check the region located at every point, which is infinite. Second, whenever a spatial object enters or leaves the sliding windows, the burst score of any region which encloses this object will change. This implies that the location of the bursty region may change as well and we need to recompute the new bursty region. With the high arrival rate of the stream, it demands an efficient strategy to update the bursty region.

In this section, we present a solution to address the \brm problem. We first introduce the \emph{\textbf{c}ontinuou\textbf{s} \textbf{b}ursty p\textbf{o}int de\textbf{t}ection} (\bpm) problem in Section~\ref{subsec:baseline}. We show that by reducing the \brm problem to the \bpm problem, for any snapshot of the stream, we convert the challenge of selecting a point from infinite points in the preferred area $A$ to selecting a \textbf{bursty point} from $O(n^2)$ disjoint regions. To address the second challenge, we present a cell-based algorithm to continuously update the \textbf{bursty point} in Section~\ref{sec:handlestream}.

\subsection{The cSPOT Problem} \label{subsec:baseline}


We next define the \bpm problem and present how to reduce the \brm problem to the \bpm problem. Firstly, we introduce some terminology that will be used to define the \bpm problem.

\begin{definition}\textbf{\textup{Rectangle Object.}}
A rectangle object, denoted with a triple $g=\langle w, \rho, t_c\rangle$, is a rectangle of size $a\times b$, where $g.w$ is its weight, $g.\rho$ is the location of its left-bottom corner, and $g.t_c$ is the creation time of $g$.
\end{definition}

Given the stream of spatial objects $\mathcal{O}$, each spatial object $o$ in $\mathcal{O}$ can be mapped to a rectangle object $g$ by using $o$ as the left-bottom corner, i.e., $g.w=o.w$, $g.\rho = o.\rho$, and $g.t_c = o.t_c$. Let $\mathcal{G}$ denote the stream of rectangle objects that are mapped from $\mathcal{O}$. Let $G(p, W)$ be the set of rectangle objects which covers point $p$ and is created in window $W$, i.e., $G(p, W) = \{g|g.t_c\in W \wedge p \in g \wedge g \in \mathcal{G}\}$. 

Next, we define the burst score of a point by following the definition of burst score of a region in Section~\ref{sec:definition} . With a slight abuse of notation, we continue to use $f(p, W)$ and $\mathcal{S}(p)$ to denote the score of a point $p$ w.r.t. the window $W$, and the burst score of $p$, respectively.

\begin{definition}\textbf{\textup{Burst Score of a Point.}}
Consider a stream of rectangle objects $\mathcal{G}$. The burst score $\mathcal{S}(p)$ of point $p$ is defined as 
$$\mathcal{S}(p) = \alpha \max (f(p, W_c) - f(p, W_p), 0) + (1-\alpha) f(r, W_c)$$
where $W_c$ and $W_p$ are the current and past windows, and for a sliding window $W$, score $f(p, W)$ is the summation of weights of rectangle objects in $G(p, W)$, i.e., $f(p, W) = \frac{\sum_{g\in G(p, W)}g.w}{|W|}$, which is the score of a point $p$ w.r.t. the sliding time window $W$.
\end{definition}

\begin{figure}[t]
  \centering
  \captionsetup{justification=centering}
  \includegraphics[width=0.5\linewidth]{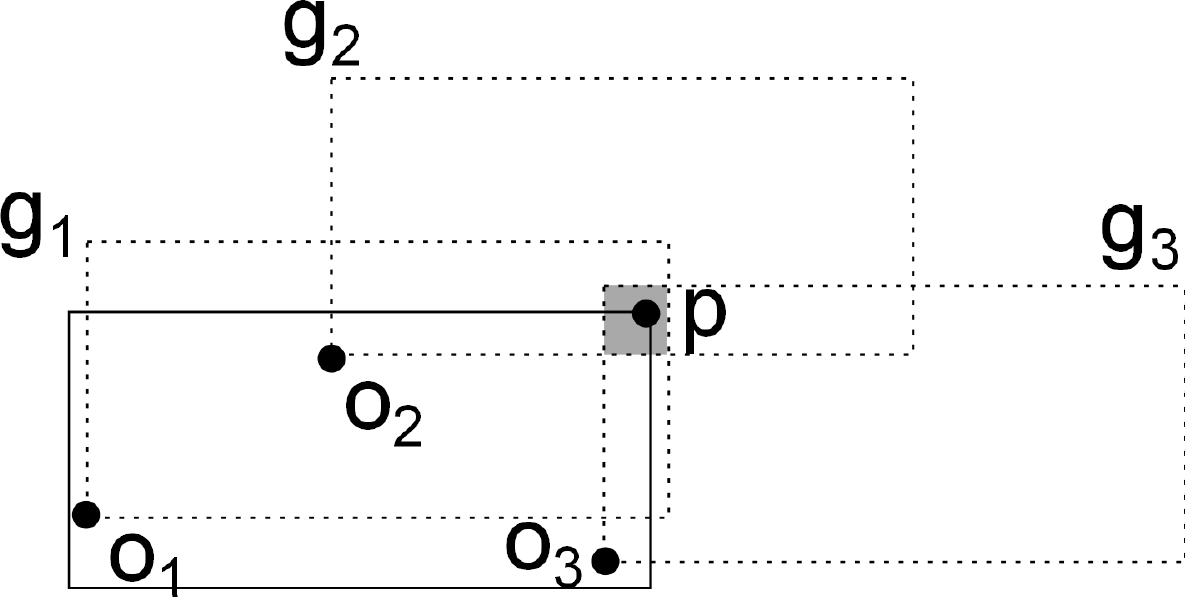}
  \caption{Reduce to cSPOT problem}\label{fig:reduce}
  \vspace{-3ex}
\end{figure}

We are now ready to formally define the \bpm problem.

\begin{definition}\label{def:burstpoint}\textbf{\textup{cSPOT Problem.}} Consider a stream of rectangle objects $\mathcal{G}$, a parameter $\alpha$, as well as the current window $W_c$ and past window $W_p$. The \textit{Continuous Bursty Point  Detection} (\bpm) problem aims to keep track of a point $p$ in the space, such that its burst score $\mathcal{S}(p)$ is maximized. A point $p$ with the maximum score is referred to as \textbf{bursty point}.
\end{definition}

In order to reduce the \brm problem to the \bpm problem, for each spatial object $o$ in the \brm problem, if $o$ is in the preferred area $A$, i.e., $o.\rho \in A$, we generate a rectangle object $g$ of size $a\times b$ with $o$ as the left-bottom corner such that $o.t_c=g.t_c$ and $g.\rho = o.\rho$. We illustrate this reduction with the example in Figure~\ref{fig:reduce}. Assume that $o_1, \dots, o_3$ are all in $A$. For each spatial object $o_i, i\in [1,3]$, a corresponding rectangle object $g_i$ is generated. We next show the relationship between the bursty region and the bursty point of the corresponding \brm and \bpm problem.

\begin{theorem} \label{the:reduce}
Let $p_m$ be a bursty point for the reduced \bpm problem given a snapshot. The rectangular region $r_m$ of size $a\times b$ whose top-right corner is located at $p_m$ is a bursty region for the original \brm problem for the snapshot.
\end{theorem}

Note that the reduction is inspired by the idea of transforming the max-enclosing rectangle problem to the \emph{rectangle intersection problem}~\cite{nandy1995unified}. The \emph{rectangle intersection problem} aims to find the most overlapped area given a set of rectangles. Since our problem has a different burst score function, the techniques designed for the rectangle intersection problem cannot be utilized to search for the bursty point at a snapshot.

We address the \brm problem by solving the corresponding \bpm problem. Observe that in the \bpm problem, the edges of the rectangle objects divide the space into many disjoint regions. Consider the example in Figure~\ref{fig:reduce}. The shaded area is one of the disjoint region which is the overlap of $g_1$, $g_2$, and $g_3$. All points in a disjoint area are covered by the same set of rectangles. Thus they have the same burst score. Next we present a theorem which justifies the reason behind the reduction.

\begin{theorem}\label{theorem:bound}
Given a snapshot of the stream of rectangle objects in the \bpm problem, there are at most $O(n^2)$ disjoint regions, where $n$ is the number of rectangle objects in windows $W_c$ and $W_p$.\cite{nandy1995unified}.
\end{theorem}

Since all points in a disjoint region have the same burst score, Theorem~\ref{theorem:bound} tells us that we only need to consider $O(n^2)$ disjoint regions, which addresses the first challenge of the \brm problem, i.e., locating the bursty region from infinite possible locations.

\begin{example}
	Consider a snapshot of the stream shown in Figure~\ref{fig:reduce}. Assume that $o_1$, $o_2$ and $o_3$ are three spatial objects in the current window $W_c$ in the \brm problem, and $o_i.w=1$ for $i\in[1,3]$. According to the reduction process, $g_1$, $g_2$ and $g_3$ are three rectangle objects in the current window in the \bpm problem, and $g_i.w=1$ for $i\in [1,3]$. Assume that $|W_c|=1$. The shaded area is the intersection of $g_1$, $g_2$ and $g_3$. Thus, any point $p$ in the shade area has the maximum burst score, i.e., $\mathcal{S}(p) = 3$. The point $p$ in the figure is a bursty point at the given snapshot. The solid line rectangle, whose top-right corner lies in $p$, is the bursty region as it encloses three spatial objects and its burst score is $3$.
\end{example}

We next present an exact solution to address the \bpm problem efficiently. Specifically, given the stream of rectangle objects, we use a grid to divide the space into cells, and maintain the upper bounds of burst score for the points in each cell. Several optimization techniques are proposed to avoid redundant recomputation. If the upper bound of any cell is larger than the score of the current bursty point, we invoke a sweep-line based algorithm to search the cell to update the location of the bursty point.  

In the rest of this section, we first introduce the sweep-line based algorithm, which finds the bursty point given a set of rectangle objects ( Section~\ref{sec:sweepline}). Then we present the cell-based lazy update strategy, which determines whether we should invoke the sweep-line algorithm to recompute the bursty point (Section~\ref{sec:handlestream}).

\begin{algorithm}[t]
\caption{\textsc{sl-cSpot} Algorithm}
\label{alg:search}
\begin{small}
\SetArgSty{textnormal}
\KwIn{A set of rectangle objects $G$}
\KwOut{A bursty point $p$}
$p=null$\;
\While{sweep-line meets an horizontal edge of a rectangle $g$}{
    $I_i, \dots, I_j \gets $ the intervals covered by $g$\;
    \For{interval $I \in \{I_i, \dots, I_j\}$}{
        Update $I.f_c$, $I.f_p$ and $I.\mathcal{S}$\;
        \If{$I.\mathcal{S} > \mathcal{S}(p)$}{
            $p\gets $ a point beneath $I$, and between the sweep-line and next horizontal edge\;
        }
    }
}
\Return{$p$}\;
\end{small}
\end{algorithm}

\subsection{Detecting Bursty Point on a Snapshot}\label{sec:sweepline}

To address the first challenge, i.e., detecting the bursty point given a snapshot of the stream, we propose a sweep-line based algorithm called \textsc{sl-cSpot} in this subsection.

\begin{figure}[t]
	\centering
	\includegraphics[width=0.7\linewidth]{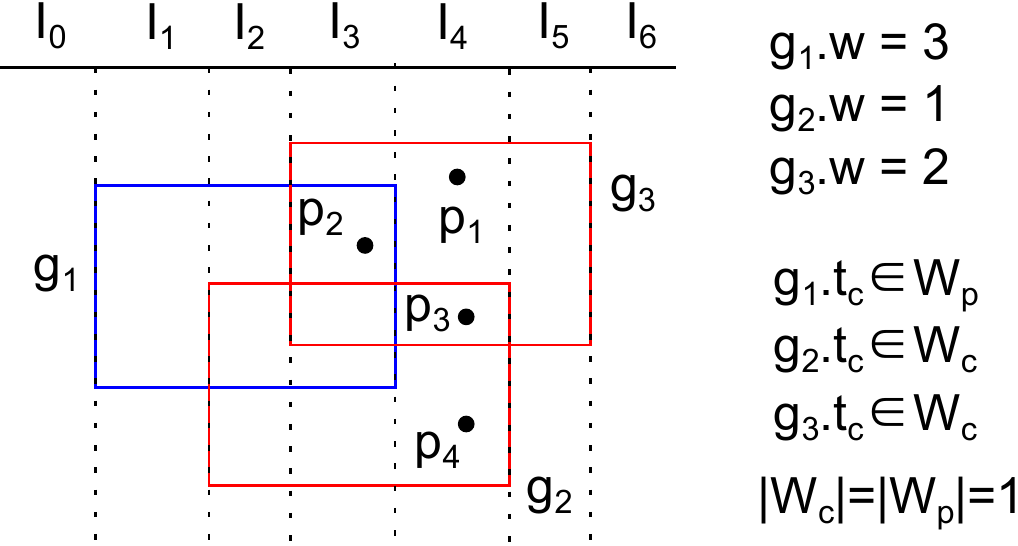}
	\caption{Illustration of 
	bursty point detection.}\label{fig:sweeplineexample}
	\vspace{-4ex}
\end{figure}

The high level idea of the \textsc{sl-cSpot} algorithm is as follows. We use a horizontal line, referred to as the sweep-line, to scan the space top-down. The sweep-line is divided into $2n + 1$ intervals at most by the vertical edges of the $n$ rectangle objects. For instance, in Figure~\ref{fig:sweeplineexample}, the vertical edges of the three rectangles divide the sweep-line into 7 intervals, $\{I_0, \dots, I_6\}$. For each interval $I$, we use $I.f_c$ and $I.f_p$ to denote the score w.r.t. the current and past windows, respectively for the points on the interval $I$. We use $I.\mathcal{S}$ to denote the burst score of such points. 
For any interval $I_i$, the set of rectangles which can cover interval $I_i$ changes when the sweep line meets the top or bottom edge of a rectangle which can cover $I_i$, and its burst score $I_i.\mathcal{S}$ is updated accordingly. A point with the maximum burst score during the sweeping process is returned as the bursty points.

We next illustrate the algorithm with an example. Figure~\ref{fig:sweeplineexample} shows a snapshot of the stream. Rectangle $g_1$ is in the past window $W_p$ (marked in blue), while $g_2$ and $g_3$ are in the current window $W_c$ (marked in red). 
As shown in Figure~\ref{fig:sweeplineexample}, when the sweep-line meets the top edge of $g_3$, any point, such as $p_1$, which is beneath the overlapped intervals  $I_3$, $I_4$ and $I_5$ and above the next horizontal line, will be covered by $g_3$. Since $g_3$ is in the current window, the score of $p_1$ w.r.t. $W_c$ will be increased by $\frac{g_3.w}{|W_c|} = 2$, resulting in an increase of its burst score. We set $I_i.f_c = 2$ and $I_i.f_p = 0$ for $i\in [3,5]$, and thus $I_i.\mathcal{S} = 0.5 \cdot \max(I_i.f_c - I_i.f_p, 0) + 0.5 \cdot I_i.f_c =2$ for $i\in [3,5]$. We select $p_1$ as the current bursty point. Then the sweep-line meets the top edge of $g_1$ and $g_2$, consecutively. The two edges are processed similarly, and we have $I_4.\mathcal{S} = 3$. Thus $p_3$ is selected as the new bursty point. When the sweep-line meets the bottom edge of the rectangle $g_3$, any point, such as $p_4$, which is beneath the overlapped intervals and above the next horizontal line, will no longer be covered by $g_3$. Thus, the scores w.r.t. $W_c$ of the overlapped intervals $I_3, \dots, I_5$ are decreased. We have $I_i.f_c = 1$ for $i\in [3,4]$, and $I_5.f_c=0$. Their burst scores are updated as: $I_3.\mathcal{S} = 1 - \alpha$, $I_4.\mathcal{S} = 1$ and $I_5.\mathcal{S} = 0$. We repeat this process until the whole space is scanned. Point $p_3$ has the maximum burst score during the sweeping process. Thus $p_3$ is returned as the bursty point.

Algorithm~\ref{alg:search} outlines this procedure. It takes as input a set of rectangle objects $G$, and outputs a bursty point $p$ with the maximum burst score in the space. Result point $p$ is initialized as $null$. The algorithm uses a sweep-line to scan the space (lines 2--7). When it meets an horizontal edge of a rectangle $r$, it first locates the intervals that are covered by $r$ (line 3). Then it updates $I.\mathcal{S}$ for each interval $I$ one by one (line 5). The point $p$ is updated if any interval has a larger burst score (lines 6--7).

\vspace{1ex}\noindent\textbf{Time Complexity.} Let $n$ be the number of rectangles in the space. The sweep-line scans $2\cdot n$ edges (each rectangle has two horizontal edges). In the worse case, when the sweep-line meets an horizontal edge, $2\cdot n + 1$ intervals are all affected. As a result, the time complexity of Algorithm~\ref{alg:search} is $O(n^2)$.

\subsection{Handling the Stream}\label{sec:handlestream}
We have presented Algorithm \textsc{sl-cSpot} to detect a bursty point given a snapshot of the stream. But how to continuously detect the bursty point? Recall that the burst score of a point is determined by the set of rectangle objects that cover it. The bursty point is likely to change when a rectangle object enters or leaves the sliding windows. Specifically, any of the following events may change the bursty point: (1) a new rectangle object enters the current window, (2) an existing rectangle object leaves the current window and enters the past window, and (3) an existing rectangle object leaves the past window. We refer to these three events as a \textit{new event}, a \textit{grown event}, and an \textit{expired event}, respectively. We use a tuple $e=\langle g, l \rangle$ to denote an event, where $g$ is the rectangle object, and $l$ is one status from $\{New, Grown, Expired\}$ to indicate the type of the event.

Intuitively, a na\"{i}ve idea is whenever an event happens, we invoke Algorithm~\ref{alg:search} to detect a bursty point on the snapshot of the stream. However, this idea does not address the \bpm problem efficiently. First, it is not necessary to search the whole space. When an event happens, it only affects the burst score of the points inside the rectangle object of the event. Second, frequent recomputation of the bursty point is computationally expensive. To address the two issues, we next present a cell-based algorithm called \textit{Cell}-\textsc{cSpot}.

\begin{algorithm}[t]
	\caption{\textit{Cell}-\textsc{cSpot} Algorithm}
	\label{alg:filter}
	\begin{small}
		\SetArgSty{textnormal}
		\KwIn{An event $e=\langle g, l\rangle$}
		\KwOut{A bursty point}
		$C_g \gets $ cells that are overlapped with $g$\;
		\For{$c \in C_g$}{
			Update $U(c)$ using Eqn~\ref{euqation:coldUB}, \ref{equ:hubupdate}, and status of $c.p$ using Lemma ~\ref{lemma:pointcorrect}\;
		}
		
		$c\gets \arg\max U(c)$\;
		\While{$c.p$ is \textit{invalid}}{
			$c.p \gets $ \FuncSty{\textsc{sl-cSpot}($c$)}\;
			$U_d(c) = \mathcal{S}(c.p)$\;
			$c\gets \arg\max U(c)$\;
		}
		\Return $c.p$
	\end{small}
\end{algorithm}

\subsubsection{Cell-based Lazy Update}\label{sec:cellalgorithm}

An event only affects the burst scores of the points inside the rectangle of the event. This locality property motivates us to divide the space into \emph{cells}, and develop approaches to handle the cells that are affected by an event. We first define the \emph{grid} that we use as follows.

\begin{definition}\label{def:gridcell}\textbf{\textup{Grid and Cell.}} We consider a grid as a set of vertical and horizontal lines defined by
	$x = i \cdot b, y = i \cdot a$ for all integers $i \in [-\infty, +\infty]$.
	For each cell $c$, we maintain a list of rectangle objects which overlap with the cell over the two sliding time windows $W_c$ and $W_p$, denoted by $c.G$.
\end{definition}

We have the following lemma based on obvious observations.

\begin{lemma}\label{lemma:affect4cells}
	A rectangle object of size $a\times b$ overlaps with at most four cells of the grid in Definition~\ref{def:gridcell}.
\end{lemma}

For each cell in the grid, we maintain a burst score upper bound for the points inside the cell (to be discussed in Section~\ref{sec:upperbound}). When an event happens, the corresponding rectangle can only affect at most four cells. Instead of searching the affected cells immediately after an event happens, we propose a \textit{lazy update strategy} by utilizing the maintained upper bound: Whenever an event happens, we first update the upper bounds of the affected cells. Then, we invoke Algorithm~\ref{alg:search} to search the cells iteratively in the descending order of their upper bounds. In each iteration, we always search the cell with the maximum upper bound. We terminate the process when there is no upper bound larger than the current maximum burst score. Hence, when an event happens, if the upper bounds of the affected cells are less than the current maximum burst score, these cells will not be searched. Thus the lazy update strategy significantly reduces the number of times that Algorithm~\ref{alg:search} is invoked to search affected cells.

In addition, to reuse the result of Algorithm~\ref{alg:search} from previous iterations, we record the point returned by Algorithm~\ref{alg:search} for each cell which is called \emph{candidate point}. The status of each candidate point is either \textit{valid} or \textit{invalid}. If the candidate point of a cell is guaranteed to have the maximum burst score in the cell, its status is \textit{valid}. On the other hand, the status is set to invalid if it is unknown whether the candidate point has the maximum burst score. We do not need to invoke Algorithm~\ref{alg:search} to search a cell if its candidate point is valid. By exploiting the candidate points, we can further avoid searching in some cells (discussed in Section\ref{sec:pointcache}).

Algorithm~\ref{alg:filter} presents an overview of our algorithm called \textit{Cell}-\textsc{cSpot} (\textit{cell}-based \textsc{cSpot}). It takes as input an event $e=\langle g, l\rangle$, and reports a bursty point in the space. The algorithm first locates the set $C_g$ of cells that overlap with $g$ (line 1). Then for each cell $c$ in $C_g$, it updates its upper bound  based on Equations~\ref{euqation:coldUB}, and \ref{equ:hubupdate} (to be introduced in Section~\ref{sec:upperbound}), and determine the status of the candidate point $c.p$ based on Lemma~\ref{lemma:pointcorrect} (to be introduced in Section~\ref{sec:pointcache}) (line 3). Then it accesses the cells in descending order of their upper bounds $U(c)$ iteratively (lines 4--8). In each iteration, if the candidate point $c.p$ is \textit{invalid}, we invoke Algorithm~\ref{alg:search} to search the cell and update $c.p$ (line 6) and the upper bound (line 7). Otherwise $c.p$ is \textit{valid}, and this indicates that $c.p$ has the maximum burst score in cell $c$ and $c$ has the maximum burst score as there is no cell whose upper bound is larger than the current maximum burst score. Therefore we terminate the process and report point $c.p$ as the result.

\vspace{0.5ex}\stitle{Time Complexity.} According to Lemma~\ref{lemma:affect4cells}, at most four cells are affected by an event rectangle $g$. Thus, it takes $O(1)$ time to update the upper bounds and candidate points. A cell will not be searched unless it is overlapped with a rectangle object. Thus, $O(1)$ cells are searched in processing a rectangle object. In our implementation, we use a heap to maintain the cells based on their upper bounds. Let $|c_{max}|$ be the maximum number of rectangle objects in a cell. Let $n$ be the number of rectangle objects created in $W_c$ and $W_p$. It takes $O(\log n)$ time to get the cell $c$ and $O(|c_{max}|^2)$ time to search the cell. Putting these together, the complexity of Algorithm~\ref{alg:filter} is $O(|c_{max}|^2+\log n)$.

\stitle{Space Complexity.} Each rectangle object is stored in at most four cells. Thus, the space cost of Algorithm~\ref{alg:filter} is $O(n)$.

\subsubsection{Upper Bound Estimation}\label{sec:upperbound}
Next, we present the details about estimating the upper bound for a cell.

\smallskip\noindent\textbf{Static Upper Bound.} We first consider a simple strategy to estimate an upper bound for a cell. According to the definition of the burst score, rectangle objects in the current window have a positive impact on the burst score, while the rectangle objects in the past window have a non-positive impact. Hence, we can estimate an upper bound burst score for a cell by only utilizing the objects in the current window. We refer to this upper bound as \emph{static upper bound}.

\begin{definition}\textbf{\textup{Static Upper Bound.}}
For a cell $c$, its static upper bound is computed as follows:
\begin{equation}\label{euqation:coldUB}
U_{s}(c) = \sum_{g\in c.G \wedge g.t_c \in W_c}\frac{g.w}{|W_c|}
\end{equation}
where $c.G$ is a set of rectangle objects overlapped with $c$.
\end{definition}

Next, we show the correctness of the static upper bound.

\begin{lemma}\label{lemma:staticcorrect}
For any point $p$ in a cell $c$, we have
$\mathcal{S}(p) \leq U_{s}(c)$.
\end{lemma}

\begin{figure}[t]
	\centering
	\includegraphics[width=0.75\linewidth]{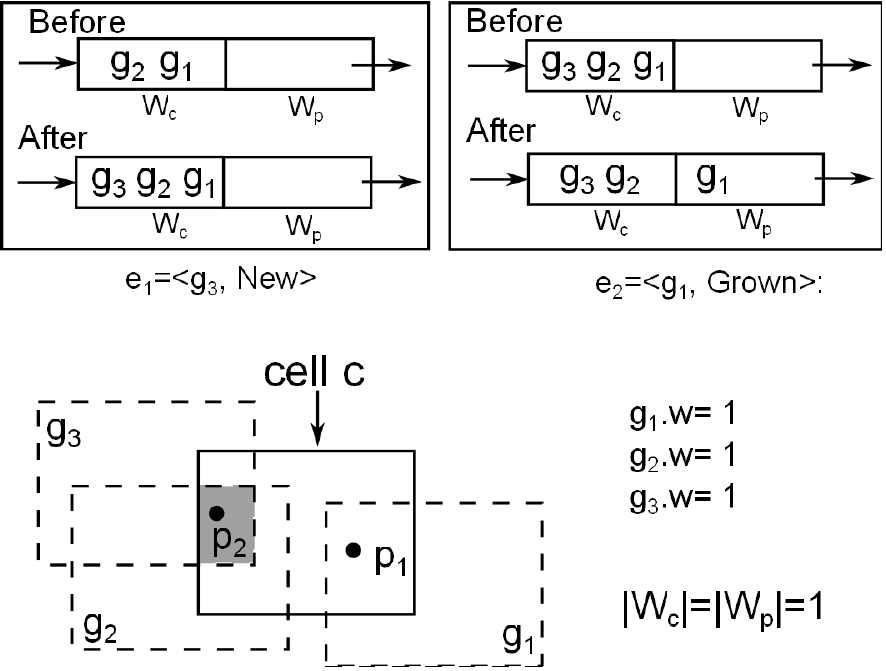}
	\caption{Cell upper bound.}\label{fig:bound}
	\vspace{-3ex}
\end{figure}

\begin{example}\label{example:loosebound}
Consider the example shown in Figure~\ref{fig:bound}. The solid-line rectangle is a cell in the grid. After event $e_1$ happens, there are three new rectangle objects overlapped with the cell $c$. The static upper bound of cell $c$ is $U_{s}(c) = 3.$
\end{example}

\vspace{1ex}\noindent\textbf{Dynamic Upper Bound.}
Next, instead of just using objects in the current window, we introduce another way to estimate the upper bound by using both the event and information from the previous computation. Specifically, when an event happens, we dynamically update the upper bound computed from previous upper bound. We refer to such upper bound as \emph{dynamic upper bound}.

Let $p_m$ be the point with the maximum burst score in cell $c$ at a snapshot $i$ when event $e_i$ arrives. Apparently $\mathcal{S}(p_m)$ is an upper bound burst score for cell $c$ at snapshot $i$. Thus, whenever we search a cell $c$ with Algorithm~\ref{alg:search} on a snapshot $i$, the dynamic upper bound $U_{d}^i(c)$ can be set as $U_d^i(c) = \mathcal{S}(p_m)$.

Let $U_d^i(c)$ be the upper bound of cell $c$ on snapshot $i$ when event $e_i$ arrives, and $U_d^{i+1}(c)$ be the upper bound when $e_{i+1}$ arrives. Let $g$ be the corresponding rectangle object of $e_{i+1}$, i.e., $e_{i+1}=\langle g, l\rangle$.
Then we have

\begin{equation}\label{equ:hubupdate}
U_{d}^{i+1}(c)=\begin{cases}
U_{d}^i(c) + \frac{g.w}{|W_c|} &\textup{$e_{i+1}.l$ is New,} \\
U_{d}^i(c) &\textup{$e_{i+1}.l$ is Grown,} \\
U_{d}^i(c) + \alpha  \frac{g.w}{|W_p|} &\textup{$e_{i+1}.l$ is Expired}
\end{cases}
\end{equation}

We next show the correctness of the dynamic upper bound with the following lemma.

\begin{lemma}\label{lemma:hubcorrect}
Consider a cell $c$. For any point $p$ in $c$, we have $\mathcal{S}(p) \leq U_{d}(c)$ after $e$ happens.
\end{lemma}

\begin{example}\label{example:dynUB}
Consider the example shown in Figure~\ref{fig:bound}. We first consider an event $e_1=\langle g_3, New\rangle$, i.e., a new rectangle object enters the current window. Assume before $e_1$ happens, we have searched the cell and the point $p_1$ has the maximum burst score in $c$. The dynamic upper bound is set as $U_d^0(c) = 1$. After $e_1$ happens, we update the dynamic upper bound as $U_d^1(c) = U_d^0(c)+\frac{g_3.w}{|W_c|}=2$. Then we consider an event $e_2=\langle g_1, Grown\rangle$, i.e., and existing rectangle object $g_1$ exits the current window and enters the past window. According to Eqn~\ref{equ:hubupdate}, the dynamic upper bound remains the same, i.e., $U_d^2(c) = 2$, since $p_2$ remains to have the maximum burst score in cell $c$.
\end{example}

We have presented the static upper bound and the dynamic upper bound. We now combine them for a tighter upper bound.

\begin{definition}\textbf{\textup{Upper bound for cell.}}
For a cell $c$, we define its upper bound $U(c)$ as $U(c) =\min(U_s(c), U_d(c)).$
\end{definition}

\subsubsection{Candidate Point Maintenance}\label{sec:pointcache}


An expensive operation of Algorithm~\ref{alg:filter} is to invoke Algorithm~\ref{alg:search} to find a point with the maximum burst score for a cell. To reuse the computation, for each cell $c$, we maintain a candidate point, denoted by $c.p$, to record the point returned by Algorithm~\ref{alg:search}. The candidate point has two possible status as introduced in Section~\ref{sec:cellalgorithm}. We next present Lemma~\ref{lemma:pointcorrect}, which is employed to determine the status of a candidate point.


\begin{lemma}\label{lemma:pointcorrect}
Let $c.p$ be a point with the maximum burst score in cell $c$ currently. Consider an event $e=\langle g, l\rangle$. After $e$ happens, if either (1) $e$ is either new or expired, $g$ can cover $c.p$, and $f(c.p, W_c) - f(c.p, W_p) > 0$, or (2) $e$ is grown object and $g$ cannot cover $c.p$, then the point $c.p$ still has the maximum burst score.
\end{lemma}

We determine the status of a candidate point based on Lemma~\ref{lemma:pointcorrect}. Consider a cell $c$ and an event $e$ which can affect $c$. If $c.p$ is valid and the conditions in Lemma~\ref{lemma:pointcorrect} hold, then $c.p$ remains to be valid. Otherwise, $c.p$ is invalid after $e$ happens.

\begin{example}
Reconsider the example shown in Figure~\ref{fig:bound}. We consider the event $e_1=\langle g_3, New\rangle$, where a new rectangle $g_3$ arrives. Before $e_1$ happens, assume that we have invoked Algorithm~\ref{alg:search} to search the cell and $p_1$ is the point with the maximum burst score. When $e_1$ happens, since $e_1$ is new and $g_3$ cannot cover $p_1$, $p_1$ is invalid after $e_1$ happens. In fact, points in the shaded area have the maximum burst score after $e_1$ happens.
\end{example}

\vspace{-1ex}\section{Approximate Solutions}\label{sec:approximate}
Although our exact solution can continuously detect the bursty region efficiently in real time, we observe that its runtime performance degrades when the number of spatial objects created in time windows $W_c$ and $W_p$ increases significantly (e.g., the sliding windows get larger, the region size gets larger, or the arrival rate of the spatial objects increases).
Since a slight imprecision is acceptable in most cases in real life, to tackle this challenge, we propose two algorithms to solve the \brm problem approximately. We prove that the burst score of the region returned by our proposed approximate algorithms is always bounded by a ratio $\frac{1-\alpha}{4}$ compared to the exact result.


\subsection{A Grid-based Solution}
The key idea behind our \textit{grid-based approximate solution} is as follows: We use a grid to divide the space into cells of size $a\times b$. Each cell is a candidate region. By maintaining the burst score for each cell, we continuously report the cell with the maximum burst score to users as an approximation to the bursty region. A nice feature of this idea is that it is intuitive while it has performance guarantees.

Algorithm~\ref{alg:gridupdate} outlines our proposed algorithm called \textsc{gap-surge} (\textbf{G}rid-based \textbf{AP}proximate \brm). Here we abuse the notation $e=\langle o, l\rangle$ to denote an event of spatial object $o$ enters or leaves the sliding windows. It first locates the cell that the spatial object $o$ lies in (line 1). The burst score of the cell $c$ is updated accordingly (lines 2--5). The cell with the maximum burst score is returned as an approximate result (line 6).

\begin{algorithm}[t]
\caption{\textsc{gap-surge} Algorithm}
\label{alg:gridupdate}
\begin{small}
\SetArgSty{textnormal}
\KwIn{An event $e=\langle o, l \rangle$} 
\KwOut{A cell $c$ }
$c_{i, j} \leftarrow $ the cell $o$ lies in\;
\lIf{$e$ is new}{	$c_{i,j}.f_c += \frac{o.w}{|W_c|}$}
\lElseIf{$e$ is grown}{	$c_{i,j}.f_c -= \frac{o.w}{|W_c|}, c_{i,j}.f_p += \frac{o.w}{|W_p|}$}
\lElse{	$c_{i,j}.f_p -= \frac{o.w}{|W_c|}$}
$c_{i,j}.\mathcal{S} = \max(c_{i,j}.f_c - c_{i,j}.f_p, 0) + c_{i.j}.f_c$\;
$c \gets \arg\max c.\mathcal{S}$\;
\Return{$c$}
\end{small}
\end{algorithm}

Before we show that the region returned by Algorithm~\ref{alg:gridupdate} has a burst score with an approximation guarantee, we present some interesting properties of the burst score function.

\begin{lemma}\label{lemma:scorefunction2}
For any two region $r_1$ and $r_2$, $r_1\subseteq r_2$, we have $\mathcal{S}(r_2) \geq (1-\alpha)\mathcal{S}(r_1)$.
\end{lemma}

\begin{lemma}\label{lemma:scorefunction3}
Let $r_1$, $r_2$ be two non-overlapping regions. We have $\mathcal{S}(r_1) + \mathcal{S}(r_2) \geq \mathcal{S}(r_1\cup r_2)$.
\end{lemma}

Now we are ready to prove the approximate ratio of Algorithm~\ref{alg:gridupdate}.

\begin{theorem}\label{theorem:gridbound}
Given a snapshot of the stream, let $r$ be the region returned by Algorithm~\ref{alg:gridupdate}, and $r_{opt}$ be the bursty region returned by our exact solution. We have $\mathcal{S}(r) \geq \frac{1-\alpha}{4} \mathcal{S}({r_{opt}})$.
\end{theorem}

\begin{lemma}\label{lemma:tight}
The approximation ratio is tight.
\end{lemma}

\noindent\textbf{Time Complexity.} In Algorithm~\ref{alg:gridupdate}, it takes constant time to locate the cell and update the burst score. In our implementation, we use a heap to maintain all cells according to their burst scores. Let $n$ be the number of spatial objects created in $W_c$ and $W_p$. Since there are $O(n)$ non-empty cells, it takes $O(\log n)$ time to report the cell with the maximum burst score.

\subsection{A Multi-Grid-Based Solution}

The burst score of the region returned by Algorithm~\ref{alg:gridupdate} is highly dependent on the position of the grid. In this subsection, we adopt multiple grids to further improve the result quality.

In the grid-based solution, we use a grid defined by lines
\begin{equation*}
\text{Grid 1: }  x = i \cdot b, y = i \cdot a
\end{equation*}
for all integers $i \in [-\infty, +\infty]$. By shifting the grid, we generate three additional grids for all integers $i\in [-\infty, +\infty]$:
\begin{equation*}
  \begin{split}
    \text{Grid 2: }&  x = 0.5b + i \cdot b, y = i \cdot a,\\
    \text{Grid 3: }&  x = b + i \cdot b, y = 0.5a + i \cdot a,\\
    \text{Grid 4: }&  x = 0.5 b + i \cdot b, y = 0.5a + i \cdot a,
  \end{split}
\end{equation*}

The multi-grid-based solution (called the \textsc{mGap-surge} algorithm) invokes Algorithm~\ref{alg:gridupdate} four times by using the four different grids. Among the four returned regions, the one with the maximum burst score is returned to users. The pseudocode of the \textsc{mGap-surge} algorithm is reported in Algorithm~\ref{alg:4grid} in Appendix~\ref{appendix:mgap-surge}.

\begin{theorem}\label{theorem:multigrid}
  The approximate ratio of the \textsc{mGap-surge} algorithm is $\frac{1-\alpha}{4}$.
\end{theorem}

\stitle{Time Complexity.} \textsc{mGap-surge} invokes Algorithm~\ref{alg:gridupdate} four times, and its complexity is $O(\log n)$, where $n$ is the number of spatial objects created in $W_c$ and $W_p$.

\vspace{-1ex} \section{Top-k Bursty Region Detection}\label{sec:topk}
Recall that in Example~\ref{example:zika}, it is paramount to monitor regions with outbreak of diseases. Intuitively, monitoring only the most bursty region is not sufficient. In fact, it is reasonable to be interested in a small list of such bursty regions. Specifically, given the size of a region, we need to continuously monitor the top-$k$ regions of the given size with highest burst scores. In this section, we present how we can elegantly extend our proposed solutions to continuously detect top-$k$ regions with highest burst scores.
We begin by formally defining the \textit{top-$k$ bursty regions}.

\subsection{Definition}
Although at first glance it may seem that it is easy to define  \textit{top-$k$ bursty regions}, in reality it is tricky. First of all, are the top-$k$ regions allowed to overlap? It may seem that detecting $k$ non-overlapping regions is a good choice. However, the non-overlapping requirement may lead us to overlooking some highly bursty regions. Hence, it is beneficial to allow the top-$k$ bursty regions to be overlapping instead of disjoint in nature.

Next, how do we define the burst scores for two overlapped regions? For example, if a spatial object lies at the intersection of two overlapping regions, which region's burst score should it contribute to? A na\"{i}ve idea is to consider it in both regions. However, this may result in $k$ regions that are highly similar to one another. To resolve this issue, we ensure that a spatial object contributes only to the burst score of at most one region.

The aforementioned considerations lead us to a greedy strategy for defining the \textit{top-$k$ bursty regions}. Specifically, given the first $i$ bursty regions, the $(i+1)$-th bursty region is the region with maximum burst score in the space but excluding all spatial objects that are already covered by the first $i$ bursty regions.

\begin{definition}\label{def:topk}\textbf{\textup{Top-$k$ Bursty Regions.}} Given $k$ rectangular regions $r_1, \dots, r_k$ such that each has a size of $a\times b$, we say $r_1, \dots, r_k$ are the top-$k$ bursty regions if and only if for any region $r$ of size $a\times b$, we have
$\mathcal{S}({r_i} \setminus {r_{[1, i-1]}}) \geq
\mathcal{S}(r \setminus {r_{[1, i-1]}})$
for $i \in [1, k]$, where ${r_{[1, i-1]}}$ is union of regions $r_1, \dots, r_{i-1}$.
\end{definition}

In order to address the top-$k$ bursty regions problem, we reduce the top-$k$ bursty regions problem to $k$ \bpm problems following the reduction in Section~\ref{subsec:baseline}. The $(i+1)$-th \bpm problem aims to detect the $(i+1)$-th bursty point from the space that excludes the set of rectangles that cover the top-$i$ bursty points.

Observe that Definition~\ref{def:topk} essentially paves the way to a greedy approach for selecting top-$k$ bursty regions.  Whenever an event happens, we can first detect a region with the maximum burst score by invoking Algorithm~\ref{alg:search}. Then we remove the spatial objects covered by the region. After that, we detect a region with the maximum burst score over the remaining objects. We repeat this process until $k$ regions are selected.

However, the na\"{i}ve strategy is inefficient as there are too many redundant computations, i.e., it is possible that we search a cell in all the $k$ reduced \bpm problems. To address the $k$ \bpm problems efficiently, we want to share the common computations among the $k$ \bpm problems. 


\subsection{Extension of the Exact Solution}
In the extension of our exact solution, for each cell $c$, we maintain $k$ upper bounds and $k$ candidate points in order to solve the $k$ \bpm problems by following the idea of Algorithm~\ref{alg:filter}. For each \bpm problem, we adopt the lazy update strategy to access the cells in descending order of their upper bounds. If the candidate point of the top cell is not valid, we search the cell by invoking Algorithm~\ref{alg:search}.

We develop two ideas of sharing computation among the $k$ \bpm problems. Firstly, if a rectangle object can cover the $i$-th bursty point, it will not be considered in the \bpm problems with order higher than $i$. For the extension, we maintain a level, denoted by $g.lvl$, for each rectangle object $g$. To select the $i$-th bursty point in response to a new event, we consider the set of rectangles $G[i:k]$ whose levels are no smaller than $i$, i.e., $G[i:k] = \{g|g.lvl \geq i\}$. When the $i$-th bursty point is selected, the levels of all the rectangles that cover the $i$-th bursty point are set as $i$, and these rectangles will not be considered by the \bpm problems with a higher order than $i$. Meanwhile, if a rectangle covers the old $i$-th bursty point, but not the new $i$-th point, its level is reset to $k$ so that it will be considered in all the $k$ \bpm problems.


Secondly, if no rectangle in a cell covers any of the $k$ detected bursty points, all the rectangles in the cell will be considered in all $k$ \bpm problems. Thus, the upper bounds and the candidate points w.r.t. the $k$ \bpm problems for the cell are the same. That is, once the upper bound and the candidate point for the cell are computed for one \bpm problem, we do not need to recompute them again for other \bpm problems.

\begin{algorithm}[t]
	\caption{\textsc{\textsc{ccs}-\textsc{Ksurge}} Algorithm}
	\label{alg:extensionExact}
	\begin{small}
		\SetArgSty{textnormal}
		\KwIn{An event $e=\langle g, l \rangle$}
		\KwOut{A bursty point}
		$g.lvl = k$, $V=\{g\}$\;
		\For{$i\in [1, k]$}{
			$p_{old} = p[i]$\;
			$C \gets $ cells that are overlapped with $V$\;
			\For{$c\in C$}{
				Update $U(c)[j]$ and $c.p[j]$ for $j\in [i, k]$\;
			}
			$c\gets \arg\max U(c)[i]$\;
			\While{$c.p[i]$ is invalid}{
				$c.p[i] \gets $ \FuncSty{\textsc{sl-cSpot}}($c$) over $G[i:]$\;
				$U_d(c)[i] = \mathcal{S}(c.p[i])$ over $G[i:]$\;
				\If{no rectangle in $c$ covers any of $p[1:k]$}{
					$c.p[1:k] = c.p[i]$, $U_d(c)[1:k] = U_d(c)[i]$\;		
				}
				$c\gets \arg\max U(c)[i]$\;
			}
			$p[i] \gets c.p$\;
			Mark $o.lvl = k$ for any $o\in G(p_{old})[i] \setminus G(p[i])[i:k]$\;
			Mark $o.lvl = i$ for any $o\in G(p[i])[i:k]$\;
			$V \gets G(p[i])[i:k] \cup G(p_{old})[i]$\;
			
		}
		\Return $p[1:k]$
	\end{small}
\end{algorithm}

Algorithm~\ref{alg:extensionExact} presents the detail of our extension. It takes as input an event $e=\langle g, l \rangle$, and output the top-$k$ bursty points, denoted by $p[1:k]$. It uses $V$ to denote the set of objects that need to be handled subsequently, and is initialized as $\{g\}$ (line 1). It then solves the $k$ \bpm problem iteratively (lines 2--17). In each \bpm problem, it first locates the set of cells affected by the objects in $V$ (line 4). For each cell $c\in C$, the upper bound $U(c)[j]$ and candidate point $c.p[j]$ w.r.t. the $j$-th \bpm problem are updated for $j\in [i, k]$ (lines 5--6). Then it accesses the cells in descending order of their upper bounds w.r.t. the $i$-th \bpm problem (lines 8--14). The upper bound and candidate point are updated as in Algorithm~\ref{alg:filter} (lines 9--10). If no rectangle in cell $c$ covers any of the $k$ detected bursty points, its $k$ upper bounds and candidate points are set to the same (lines 11-12). When a new bursty point is found, we reset the levels for the affected objects as discussed earlier (lines 15--16): The rectangles that cover the old bursty point $p_{old}$ but not the new bursty point $p[i]$ are newly visible to all the $k$ \bpm problems, while the rectangles that cover the new bursty point $p[i]$ are newly invisible to the \bpm problems with a higher order than $i$. The two types of rectangle objects comprise $V$, which will be processed in the next \bpm problem (line 17). After $k$ iterations, it returns the top-$k$ bursty points $p[1:k]$ as the result.

\vspace{1ex}\stitle{Time Complexity.} A cell is searched if its upper bound is either changed by an event or by a detected bursty point. Thus, the algorithm searches $O(1+k) = O(k)$ cells on average when processing a rectangle. The complexity of Algorithm~\ref{alg:extensionExact} is $O(|c_{max}|^2 \cdot k)$, where $|c_{max}|$ is the maximum number of objects that we search in a cell.

\subsection{Extension of the Approximate Solutions}\label{sec:kapprox}
We also extend our approximate solutions in Section~\ref{sec:approximate} to find $k$ regions with relatively high burst score.

\vspace{1ex}\stitle{Extending the \textsc{gap-surge} Algorithm.} Consider the grid-based solution.  We use a heap to maintain all cells with their burst scores. Thus, we can simply return top-$k$ cells with highest burst scores. In our implementation, we use a heap to maintain the cells. Thus, its complexity is $O(\log n)$.(Algorithm~\ref{alg:apptopk} in Appendix~\ref{appendix:algorithm}).

\vspace{1ex}\stitle{Extending the \textsc{mGap-surge} Algorithm.} We extend the multi-grid-based solution similarly. Note that one cell in a grid may overlap with at most four cells in another grid. Thus, for each grid, we maintain the top-$4k$ cells. Then we merge the $16\cdot k$ cells and return the top-$k$ non-overlapping cells. Its time complexity is $O(\log n + k)$.(Algorithm~\ref{alg:4gridtop4} in Appendix~\ref{appendix:algorithm}).


\vspace{-1ex} \section{Experimental Study}\label{sec:exp}
We investigate the performance of our proposed techniques.  All algorithms are implemented in C++ complied with GCC 4.8.2. The experiments are conducted on a machine with a 2.70GHz CPU
and 64GB of memory running Ubuntu.
\subsection{Experimental Setup}
\noindent\textbf{Datasets.} We conduct experiments on three public real-life datasets as reported in Table~\ref{tab:dataset}. \textsf{UK} consists of 1,000,000 geo-tagged tweets posted in UK.  \textsf{US} consists of 1,000,000 geo-tagged tweets posted in US and has a higher arrival rate. \textsf{Taxi}\footnote{\scriptsize \url{crawdad.org/roma/taxi/20140717}} consists of mobility traces of taxi cabs obtained from the \textsc{gps} in Roma, Italy. It contains 1,000,000 records over 5 days. For each dataset, the weight of each spatial object is randomly chosen from from [1, 100] with a uniform distribution.

\begin{table}[t]
\caption{Datasets.} \label{tab:dataset}
\vspace{-3ex}
\begin{center}
\begin{scriptsize}
\centering
\begin{tabular}[t]{| c | c | c | c | }
\hline
Datasets & \textsf{UK} & \textsf{US} & \textsf{Taxi} \\ \hline \hline
\# of Spatial Objects & 1,000,000 & 1,000,000 & 1,000,000 \\ \hline
Arrival Rate(per hour) & 5,747 & 16,802 & 18,145  \\ \hline
Range of Latitude & 139.0~150.9 & 100.1~150.4 & 41.6~42.2 \\ \hline
Range of Longitude & 171.1~181.9 & 40.2~118.8 & 12.0~12.9 \\ \hline
\end{tabular}
\end{scriptsize}
\vspace{-2ex}
\end{center}
\vspace{-3ex}
\end{table}

\vspace{1ex}\noindent\textbf{Algorithms.} We evaluate the performances of the three proposed algorithms, namely the exact method \textit{Cell}-\textsc{cSpot} (denoted by \textsf{CCS}), the grid-based approximation algorithm \textsc{gap-surge} (denoted by \textsf{GAPS}), and the multi-grid-based technique \textsc{mGap-surge} (denoted by \textsf{MGAPS}).  We denote the top-$k$ extensions of these algorithms as  \textsf{kCCS},  \textsf{kGAPS}, and  \textsf{kMGAPS}, respectively. To evaluate the usefulness of our proposed method of upper bound estimation, we compare \textsf{CCS} with an approach that only utilizes the static upper bound, denoted by \textsf{B-CCS} 
, and a baseline approach that does not use any upper bound estimation technique, denoted by \textsf{Base}. 
To the best of our knowledge, there is no existing technique that address the \brm problem. Hence we are confined to compare our proposed algorithms with \textsf{aG2}~\cite{amagata2016monitoring}, which is designed for continuously monitoring the \emph{MaxRS} problem.
In our experiments, we use a modified version of \textsf{aG2}. With a slight abuse of notation, we still use \textsf{aG2} to denote the modified \textsf{aG2}. The details of the \textsf{Base} and \textsf{aG2} are reported in Appendix~\ref{appendix:evaluatedmethods}.


\vspace{1ex}\noindent\textbf{Parameters.} By default, we set the size of the past window $W_p$ and the current window $W_c$ as 1 hour for \textsf{US} and \textsf{UK}, and 5 minutes for \textsf{Taxi}. We set the size of the query rectangle as $1/1000$ of the range of each dataset by default, denoted by $q$. We set the preferred area $A$ as the whole space. For the \textsf{aG2} algorithm, we set the size of a cell to $10q$.

\vspace{1ex} \noindent\textbf{Stream Workload.} We start the simulation when the system becomes stable, i.e., there exists an expired object from the past sliding window. We continuously run each algorithm for 1,000,000 new arriving spatial objects over the two sliding windows. The average processing time per object is reported.


\begin{figure}[t]
\centering
\vspace{-1ex}
\subfigure[Taxi]{
	\includegraphics[width=0.45\linewidth, height=2.4cm]{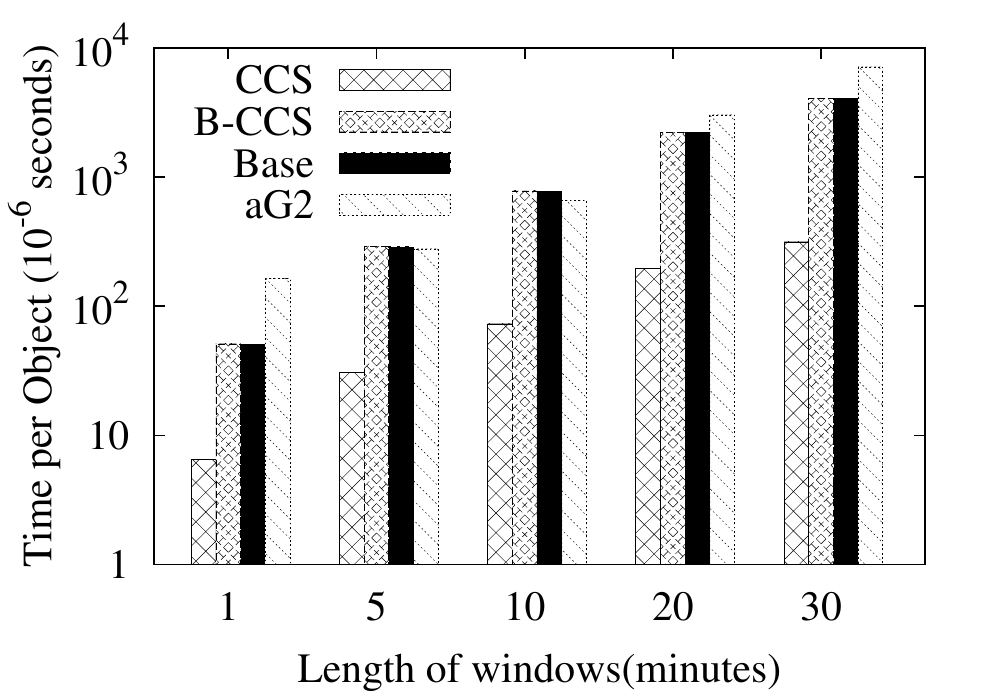}
}
\vspace{-1ex}
\subfigure[UK]{
	\includegraphics[width=0.45\linewidth, height=2.4cm]{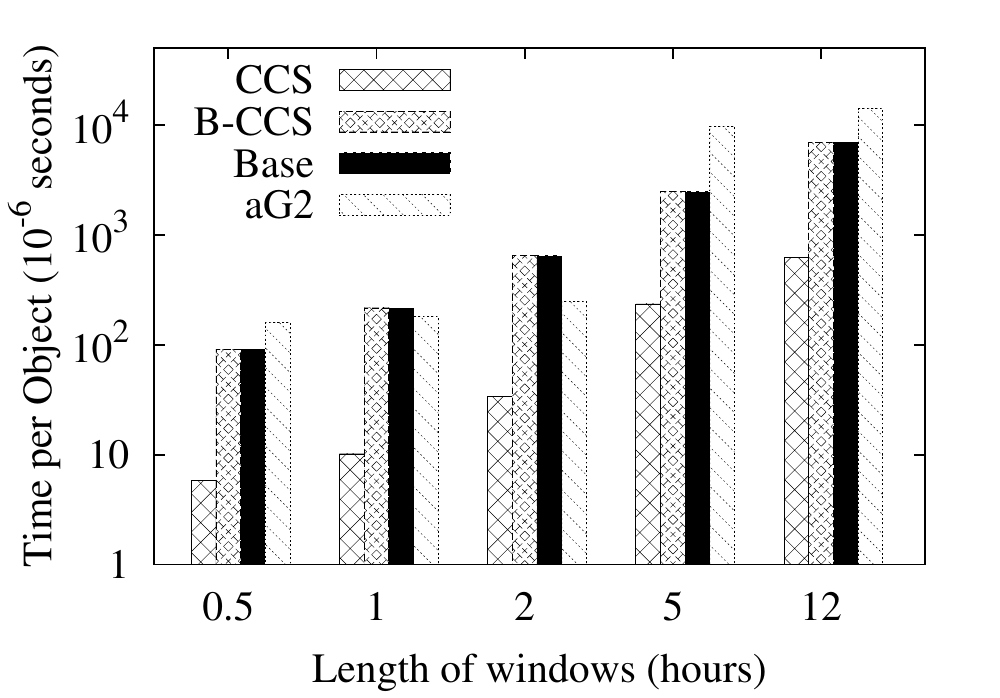}
}
\vspace{-1ex}
\subfigure[US]{
	\includegraphics[width=0.45\linewidth, height=2.4cm]{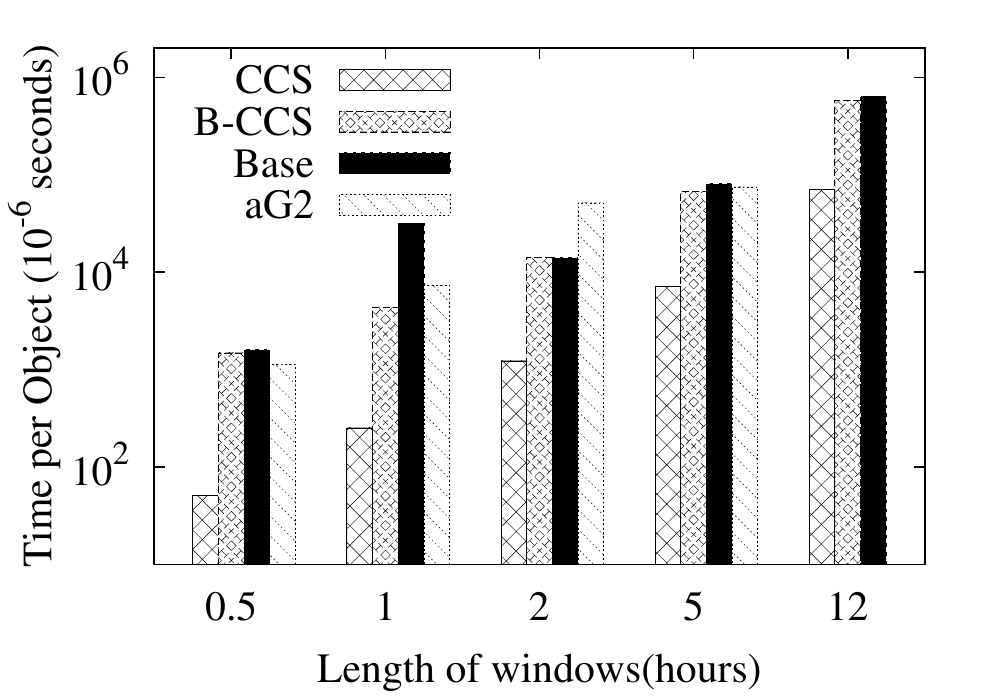}
}
\vspace{-1ex}
\subfigure[Taxi]{
\includegraphics[width=0.45\linewidth, height=2.4cm]{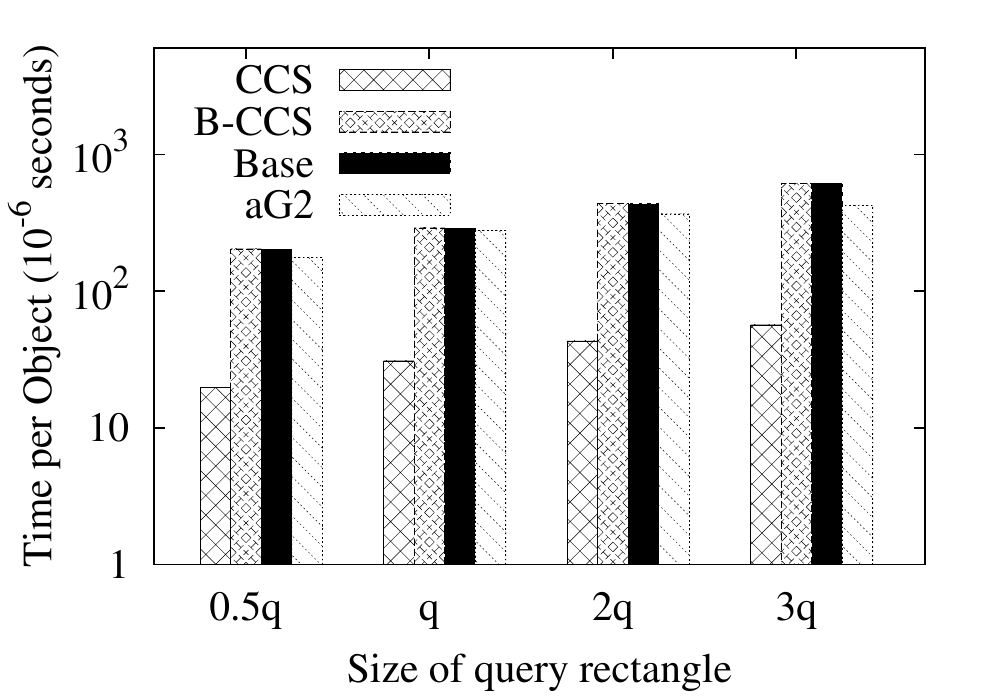}
}
\vspace{-1ex}
\subfigure[UK]{
\includegraphics[width=0.45\linewidth, height=2.4cm]{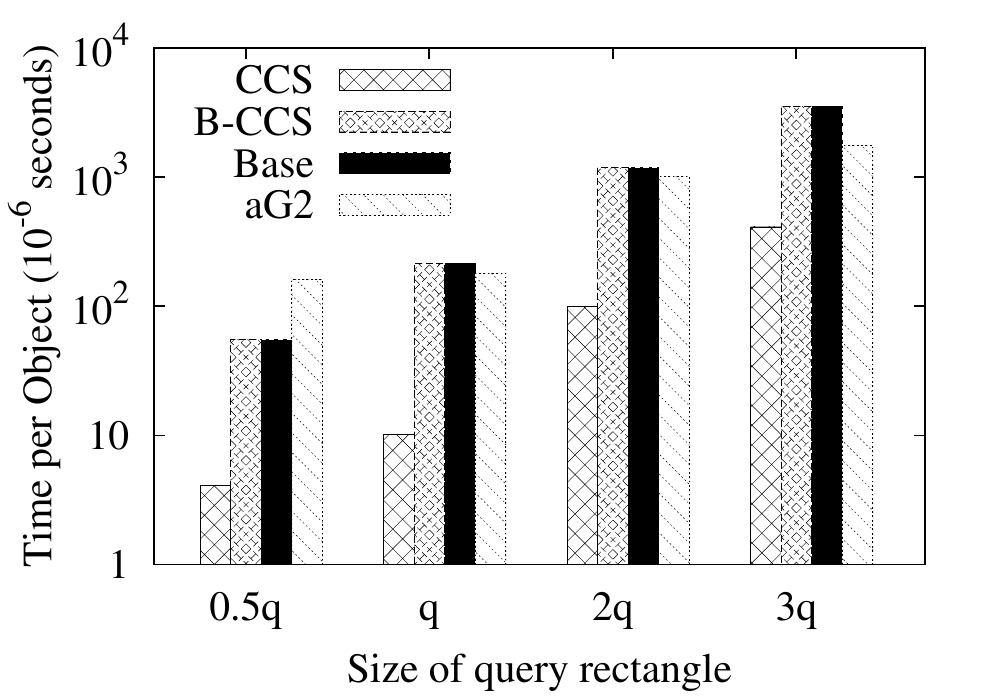}
}
\vspace{-1ex}
\subfigure[US]{
\includegraphics[width=0.45\linewidth, height=2.4cm]{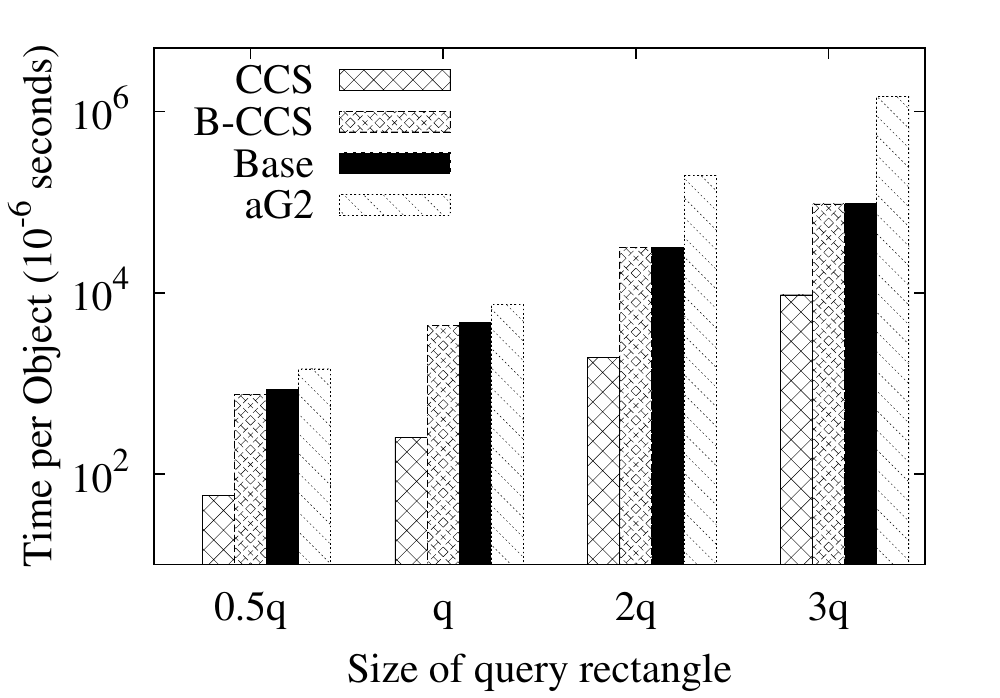}
}
\caption{Runtime of \textsf{CCS}, \textsf{B-CCS}, \textsf{Base} and \textsf{aG2}.}
\label{fig:exact}
\vspace{-2ex}\end{figure}

\subsection{Evaluation of the Exact Solution} \label{sec:exactexp}
We first evaluate the runtime performance of \textsf{CCS}, \textsf{B-CCS} and \textsf{Base} on each dataset. Then we study the usefulness of the upper bound in \textsf{CCS}.

\vspace{1ex}\stitle{Runtime Performance.} The aim of  the first set of experiments is to evaluate the efficiency of our exact solution \wrt the sliding window size and the query rectangle size. For \textsf{US} and \textsf{UK}, we vary the sliding window with the following sizes: \textit{30 minutes, 1 hour, 2 hours, 5 hours,} and \textit{12 hours}. For \textsf{Taxi}, we use the following five sizes for sliding windows: \textit{1 minute, 5 minutes, 10 minutes, 20 minutes,} and \textit{30 minutes}. We use the following four sizes for the query rectangle: $0.5q$, $q, 2q$, and $3q$. 

Figures~\ref{fig:exact}(a)--(c) report the average runtime of the three methods for processing one spatial object as we vary the size of sliding windows. Note that the y-axis is in logarithmic scale. We find that \textsf{CCS} runs efficiently and  outperforms \textsf{aG2}. For example, for \textsf{Taxi}, it takes about $3\times 10^{-4}$ seconds to process an object when the current and past windows are both set to 30 minutes, while \textsf{aG2} takes $7\times 10^{-3}$ seconds. In addition, we find that \textsf{aG2} run out of the 64 GB memory on \textsf{US} when the current window and past window are both set as 12 hours, as there are too many spatial objects in the two sliding windows. 

Moreover, we observe that the processing time per object of all algorithms increases as the size of window increases. This is due to the need to consider a larger number of spatial objects when we search for the bursty region with the increase in size of the sliding window.  Consequently, the runtime per object increases.

Figures~\ref{fig:exact}(d)--(f) report the average runtime for processing one spatial object as we vary the size of the query rectangle. Similarly, the average runtime increases as size of the rectangle increases.

\vspace{1ex}\stitle{Usefulness of Upper Bound.} Next, we evaluate the usefulness of the method for upper bound estimation in
\textsf{CCS}. In this set of experiments, we process 1,000,000 new objects and report how many rectangles trigger a search. 
The results are reported in Table~\ref{tab:sweeptime_window}. Clearly, only a small portion of rectangle messages (2\%-5\% for \textsf{Taxi}, and less than 1\% for \textsf{US} and \textsf{UK}) trigger a search in \textsf{CCS} compared with \textsf{B-CCS}. This is because \textsf{CCS} can estimate a much tighter upper bound for cells. Thus, many cells are eliminated from further checking. This also explains why \textsf{CCS} is much more efficient than \textsf{B-CCS}. 
As shown in Figure~\ref{fig:exact}, we observe that \textsf{CCS} is more efficient than the other two methods. The runtime of \textsf{CCS} is more than one order of magnitude faster than \textsf{B-CCS} and \textsf{Base}, respectively. Moreover, we observe that \textsf{B-CCS} is only marginally better than \textsf{BASE}, which indicate that  only using the static upper bound cannot effectively avoid unnecessary recomputation. This is because the static upper bound is too loose, especially when the weights of the objects are randomly chosen from 1 to 100.

\begin{table}[t]
\caption{Ratio of rectangle messages that trigger a search vs. window size for \textsf{CCS} and \textsf{B-CCS}.} \label{tab:sweeptime_window}
\vspace{-3ex}
\begin{center}
\begin{scriptsize}
\centering
\begin{tabular}[h]{| c | c | c | c | c | c | c | }
\hline
\multirow{3}{*}{Taxi}&  Window (mins) & 1  &  5  & 10  & 20  & 30  \\ \cline{2-7}
                     &   \textsf{CCS}  & 4.85\% & 3.20\% & 2.56\% &  2.13\%  & 1.95\% \\ \cline{2-7}
                     &   \textsf{B-CCS}  & 92.63\% & 78.30\% & 70.00\% &  62.07\%  & 57.90\% \\ \hline \hline
\multirow{3}{*}{UK}& Window (hours) & 0.5  &  1  & 2  & 5  & 12  \\ \cline{2-7}
                     &  \textsf{CCS}  & 0.34\% & 0.27\% & 0.23\% &  0.37\%  & 0.48\% \\ \cline{2-7}
                     &  \textsf{B-CCS}  & 37.79\% & 28.23\% & 22.76\% &  21.64\%  & 14.57\% \\ \hline \hline
\multirow{3}{*}{US}& Window (hours) & 0.5  &  1  & 2  & 5  & 12  \\ \cline{2-7}
                     &  \textsf{CCS}  & 0.60\% & 0.68\% & 0.70\% &  0.52\%  & 0.60\% \\ \cline{2-7}
                     &  \textsf{B-CCS}  & 64.21\% & 52.29\% & 35.13\% &  9.0\%  & 20.90\% \\ \hline
\end{tabular}
\end{scriptsize}
\vspace{-2ex}
\end{center}
\vspace{-1ex}
\end{table}

\begin{figure}[t]
\centering
\vspace{-1ex}
\subfigure[Taxi]{
\includegraphics[width=0.45\linewidth, height=2.3cm]{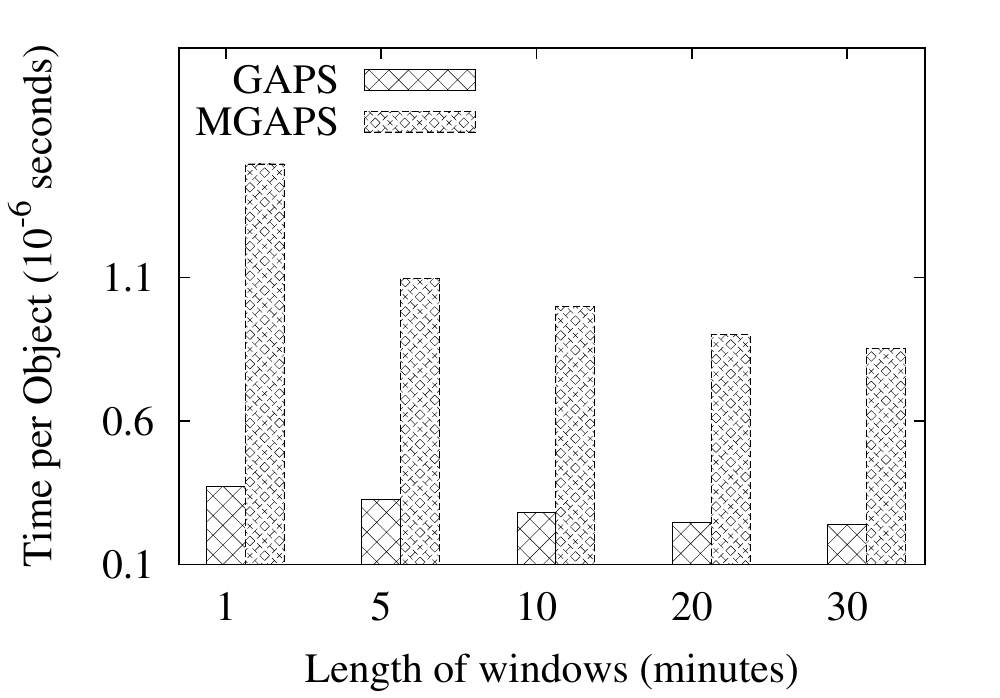}
}
\vspace{-1ex}
\subfigure[UK]{
\includegraphics[width=0.45\linewidth, height=2.4cm]{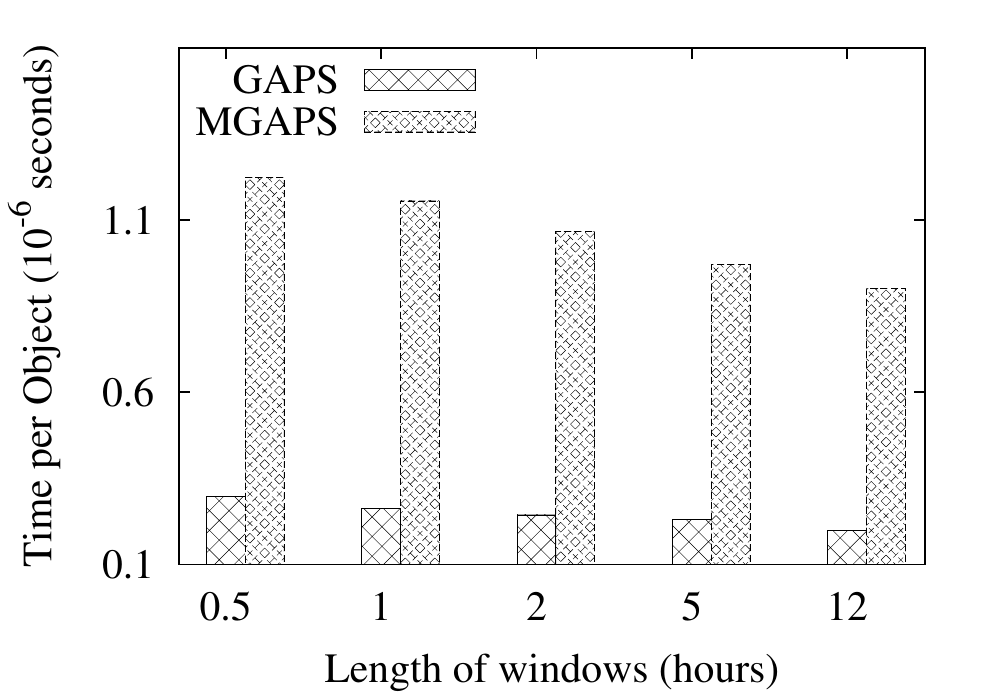}
}
\vspace{-1ex}
\subfigure[US]{
\includegraphics[width=0.45\linewidth, height=2.4cm]{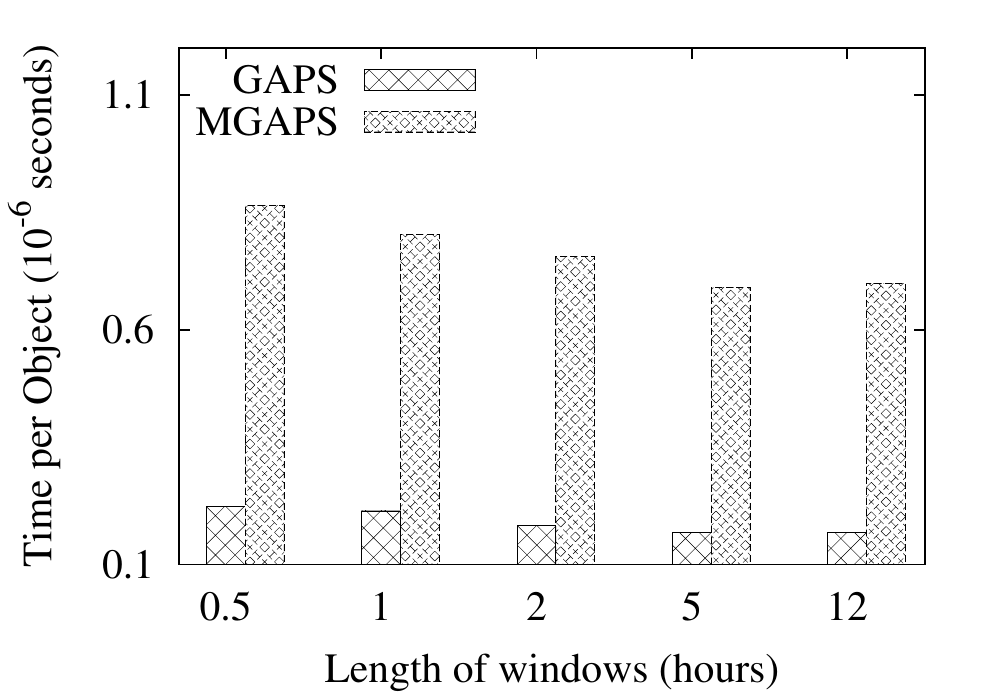}
}
\vspace{-1ex}
\subfigure[Taxi]{
\includegraphics[width=0.45\linewidth, height=2.4cm]{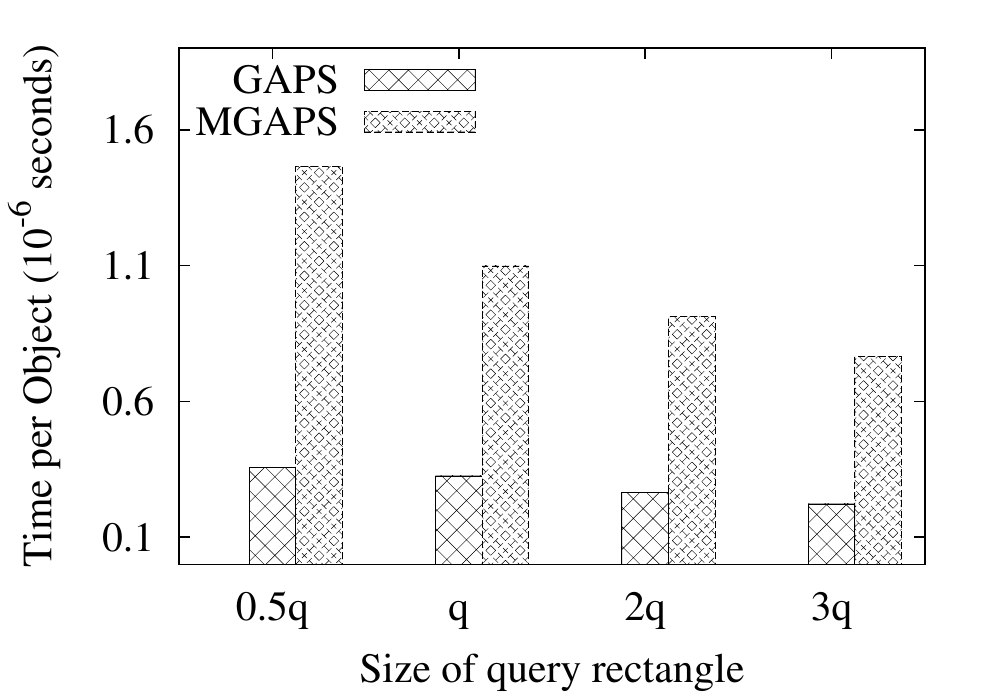}
}
\vspace{-1ex}
\subfigure[UK]{
\includegraphics[width=0.45\linewidth, height=2.4cm]{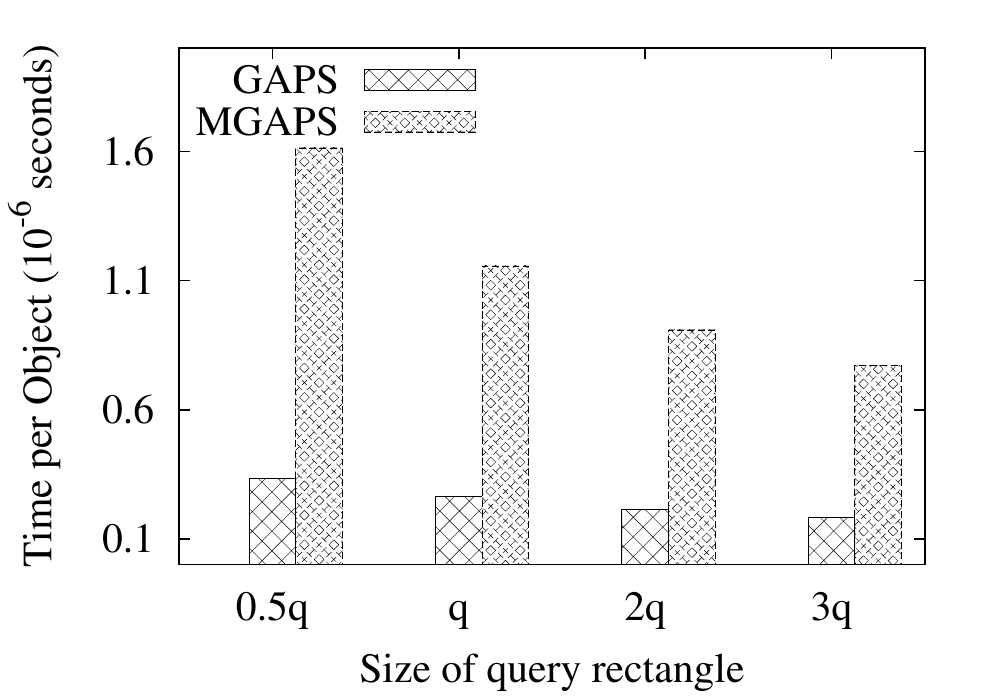}
}
\vspace{-1ex}
\subfigure[US]{
\includegraphics[width=0.45\linewidth, height=2.4cm]{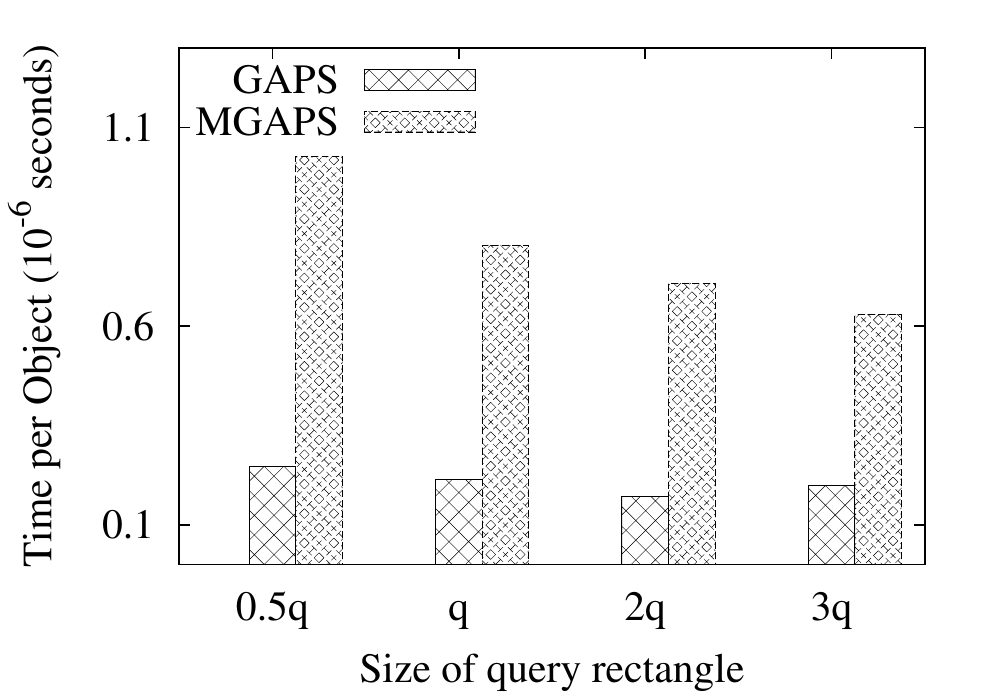}
}
\caption{Runtime performance of \textsf{GAPS} and \textsf{MGAPS}.}
\label{fig:app}
\vspace{-1ex}\end{figure}

\subsection{Evaluation of the Approximate Solutions}
The detailed results on approximation ratios are reported in Appendix~\ref{appendix:ratio}. A short summary of the results is that the burst score of the region detected by \textsf{GAPS} (resp. \textsf{MGAPS}) is about 73\% -- 92\% (resp. 85\% -- 94\%) of that of the optimal region. We next report the runtime of the approximate solutions.

\vspace{1ex}\stitle{Runtime Performance.} We evaluate the efficiency of our approximate solutions \wrt the sliding window size and the query rectangle size under the same setting as for the exact solution. Figures~\ref{fig:app} (a)--(c) report the average runtime for processing one spatial object using \textsf{GAPS} and \textsf{MGAPS} as we vary the sliding window. Figures~\ref{fig:app} (d)--(f) report the average runtime for processing one spatial object as we vary the size of the query rectangle. We find that the runtime of \textsf{MGAPS} is about 2-5 times of \textsf{GAPS}, which is expected as \textsf{MGAPS} invokes \textsf{GAPS} four times. Moreover, \textsf{GAPS} and \textsf{MGAPS} are about three orders of magnitude faster than \textsf{CCS} by comparing Figure~\ref{fig:exact} and Figure~\ref{fig:app}.

\subsection{Effect of $\alpha$}
In the definition of burst score, we use a parameter $\alpha$ to balance the significance and the burstiness. 
In this set of experiments, we evaluate the impact of the parameter $\alpha$ on the efficiency and approximation ratio of our proposed algorithms on the \textsf{US} dataset. We use 1 hour for the sliding windows and $q$ for the size of the query rectangle.

\begin{figure}[t]
\centering
\subfigure[Exact Solutions]{
\includegraphics[width=0.45\linewidth, height=2.3cm]{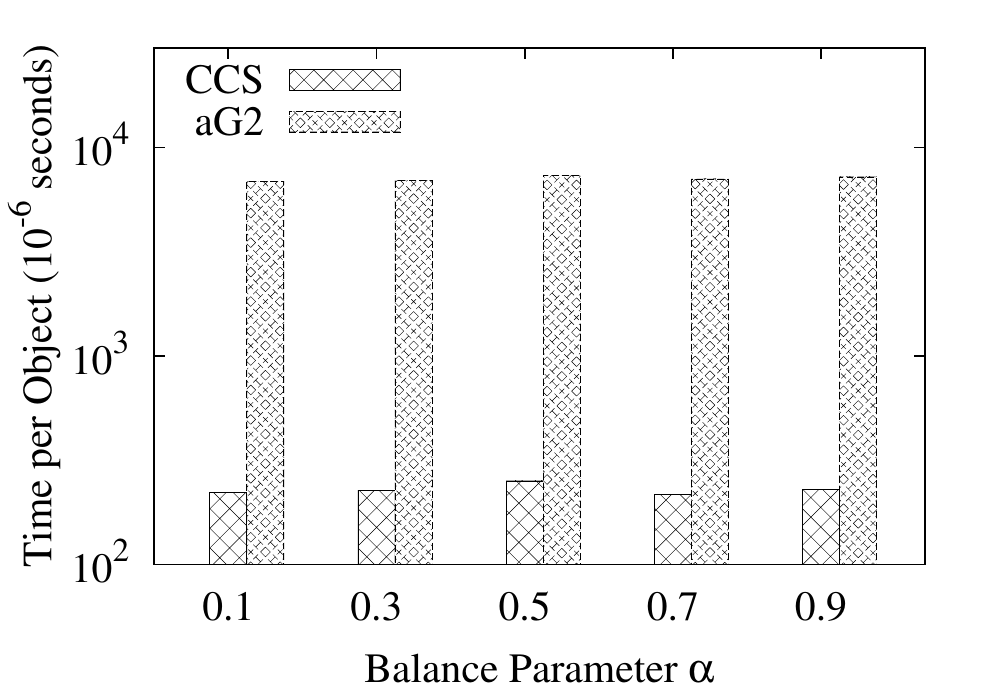}
}
\subfigure[Approximate Solutions]{
\includegraphics[width=0.45\linewidth, height=2.4cm]{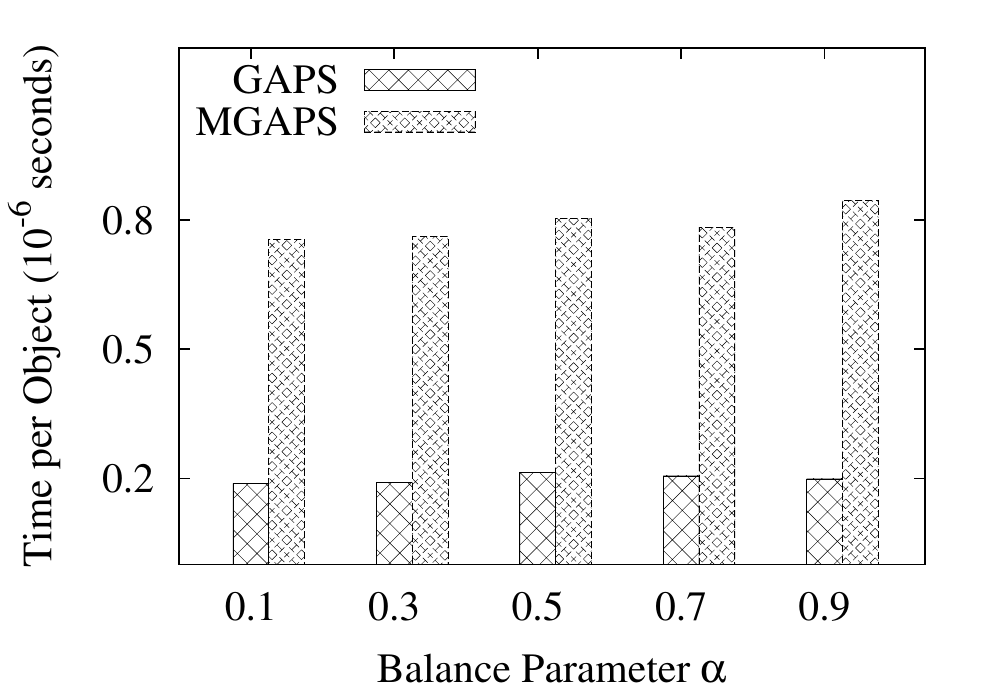}
}
\vspace{-2ex}\caption{Runtime performance w.r.t. $\alpha$ on \textsf{US}.}
\label{fig:alpha_efficiency}
\vspace{-2ex}\end{figure}

\vspace{1ex}\stitle{Impact on Runtime Performance.} We evaluate the efficiency of our exact and approximate solutions w.r.t. the balance parameter $\alpha$. Figure~\ref{fig:alpha_efficiency} reports the average runtime for processing one spatial object as we vary $\alpha$ from 0.1 to 0.9. We observe that the efficiency is hardly affected by the parameter $\alpha$ for both our exact solution and approximate solutions.

\begin{table}[t]
\caption{Approximate ratio vs. $\alpha$.} \label{tab:approxratio_alpha}
\vspace{-3ex}
\begin{center}
\begin{scriptsize}
\centering
\begin{tabular}[h]{| c | c | c | c | c | c | c | }
\hline
\multirow{3}{*}{US} & $\alpha$ & 0.1  & 0.3  & 0.5  & 0.7  & 0.9  \\ \cline{2-7}
& \textsf{GAPS} & 82.57\% & 81.76\% & 80.67\% &  77.23\% &  78.58\% \\ \cline{2-7}
& \textsf{MGAPS} & 90.50\% & 89.44\%b & 88.07\%  & 87.80\% & 86.67\% \\ \hline
\end{tabular}
\end{scriptsize}
\vspace{-1ex}
\end{center}
\vspace{-3ex}
\end{table}

\vspace{1ex}\stitle{Impact on Approximation Ratio.} In this set of experiments, we evaluate the approximate ratio of the burst scores of regions detected by \textsf{GAPS} and \textsf{MGAPS} by varying $\alpha$. The results are reported in Table~\ref{tab:approxratio_alpha}. We find that the approximate ratios of the two algorithms decrease as $\alpha$ increases. This is because their theoretical approximate ratio $\frac{1-\alpha}{4}$ decreases as $\alpha$ increases. 

\subsection{Scalability}
\begin{figure}[t]
\centering
\subfigure[\textsf{CCS}]{
\includegraphics[width=0.46\linewidth, height=2.4cm]{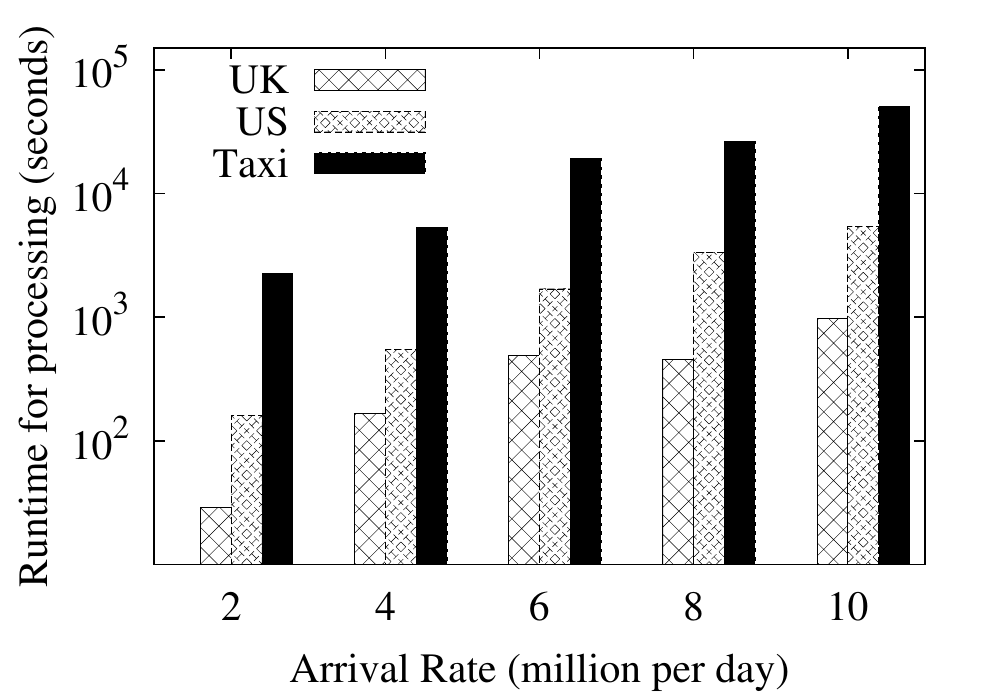}
}
\subfigure[\textsf{GAPS}]{
\includegraphics[width=0.46\linewidth, height=2.4cm]{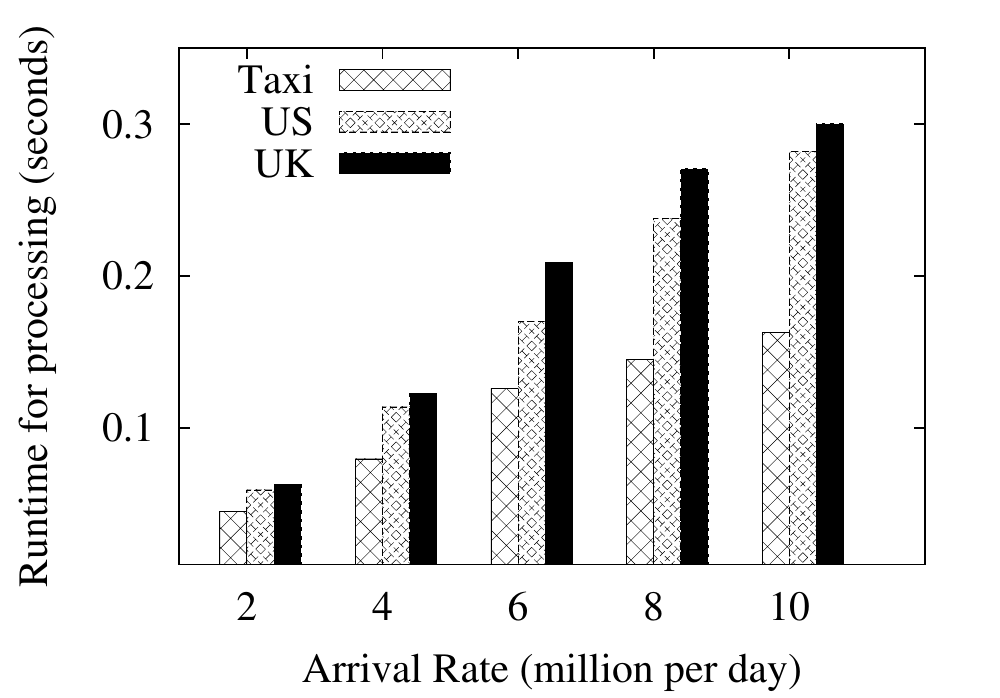}
}
\vspace{-2ex}\caption{Scalability study.}
\label{fig:arrivalrate}
\vspace{-1ex}\end{figure}

We now investigate the scalability of our proposed techniques by varying the arrival rate of the spatial objects. Specifically, we use 1 hour for both the current window and past window, and $q$ for the size of the query rectangle on all three datasets. We stretch the stream to change its arrival rate from 2 million per day to 10 million per day. For example, in \textsf{UK}, 1 million spatial objects arrived in 174 hours. Hence, we shrink the arrival time of each object to make all objects arrive in 24 hours. Then the
arrival rate of the stream is 1 millions per day. We only report the \textit{average time} for processing the objects arrived in one hour (denoted by $t_h$) of \textsf{CCS} and \textsf{GAPS} in Figure~\ref{fig:arrivalrate}. Formally, $t_h = \frac{runtime}{|\mathcal{O}|_{time}}$,
where $runtime$ is the runtime of the algorithm, and $|\mathcal{O}|_{time}$ is the total timespan of the stream.

We observe that it takes several hours for \textsf{CCS} to process the objects arrived in an hour for the \textsf{Taxi} dataset, which means that it  does not scale well and cannot handle streams with high arriving rate. On the other hand, our approximate solutions, \textsf{GAPS} and \textsf{MGAPS}, scale well with the increase in arrival rate. They can process the objects arrived in an hour within seconds. 

\begin{figure}[t]
\centering
\vspace{-1ex}
\subfigure[Taxi]{
\includegraphics[width=0.45\linewidth, height=2.4cm]{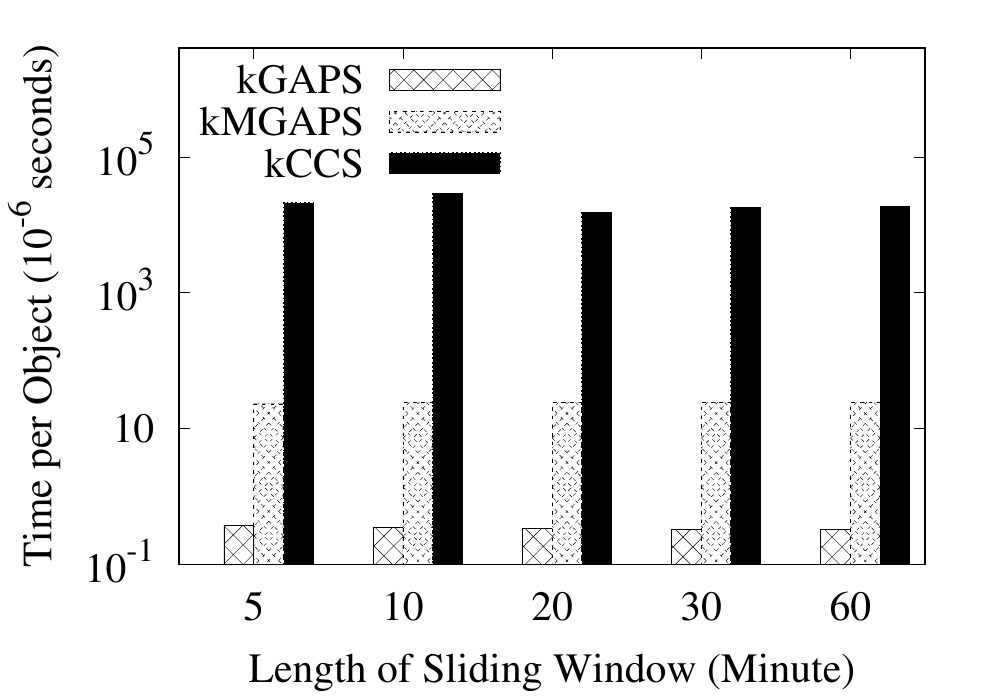}
}
\vspace{-1ex}
\subfigure[UK]{
\includegraphics[width=0.45\linewidth, height=2.4cm]{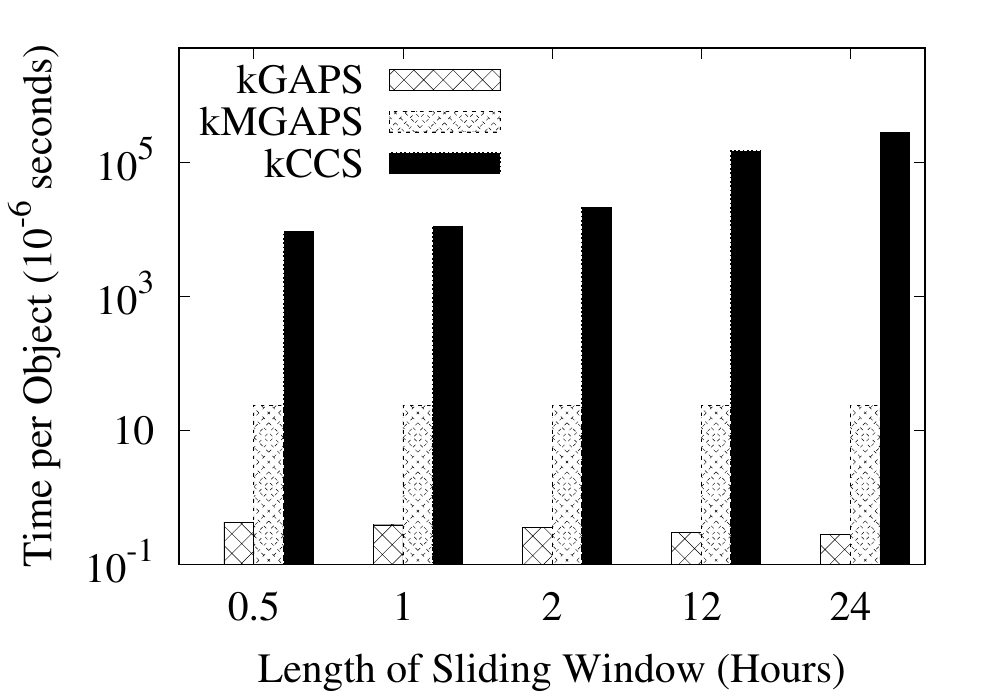}
}
\vspace{-1ex}
\subfigure[US]{
\includegraphics[width=0.45\linewidth, height=2.4cm]{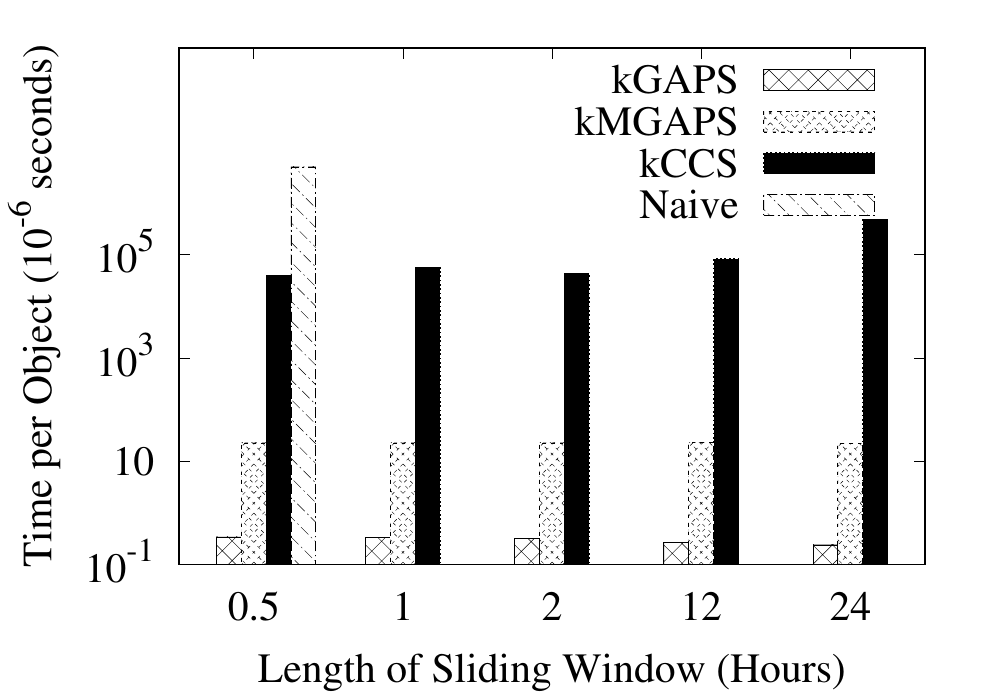}
}
\vspace{-1ex}
\subfigure[\textsf{kCCS}]{
\includegraphics[width=0.45\linewidth, height=2.4cm]{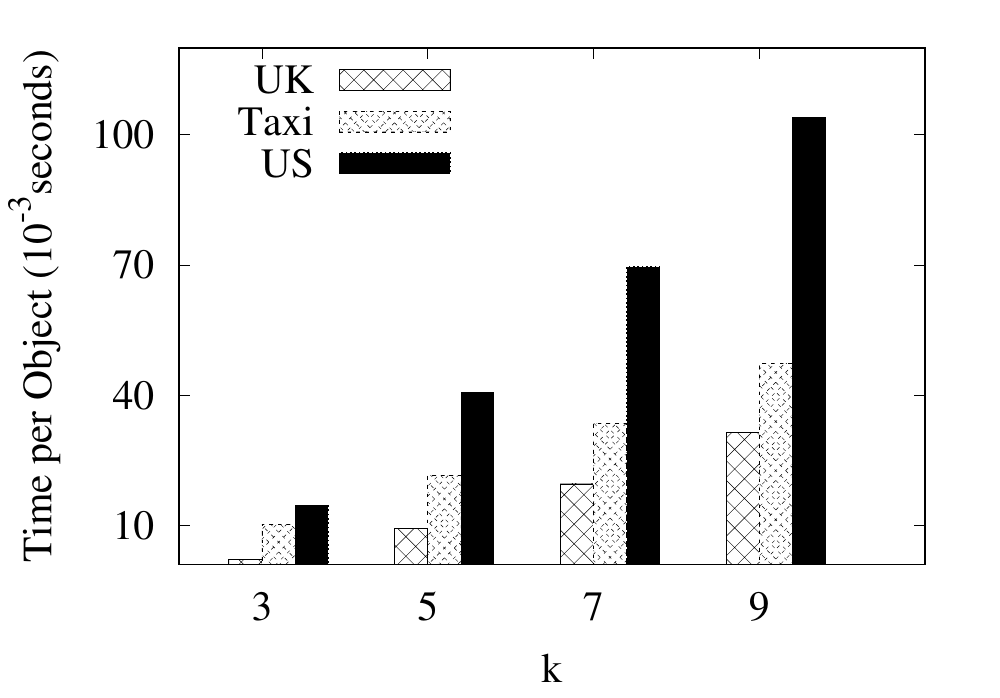}
}
\vspace{-1ex}
\subfigure[\textsf{kGAPS}]{
\includegraphics[width=0.45\linewidth, height=2.4cm]{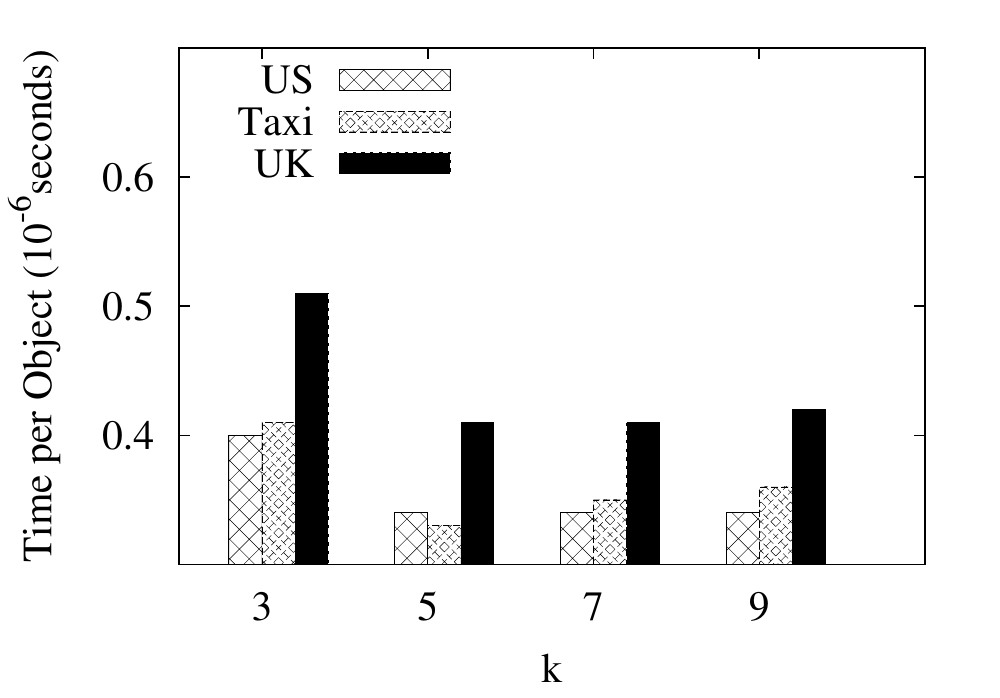}
}
\vspace{-1ex}
\subfigure[\textsf{kMGAPS}]{
\includegraphics[width=0.45\linewidth, height=2.4cm]{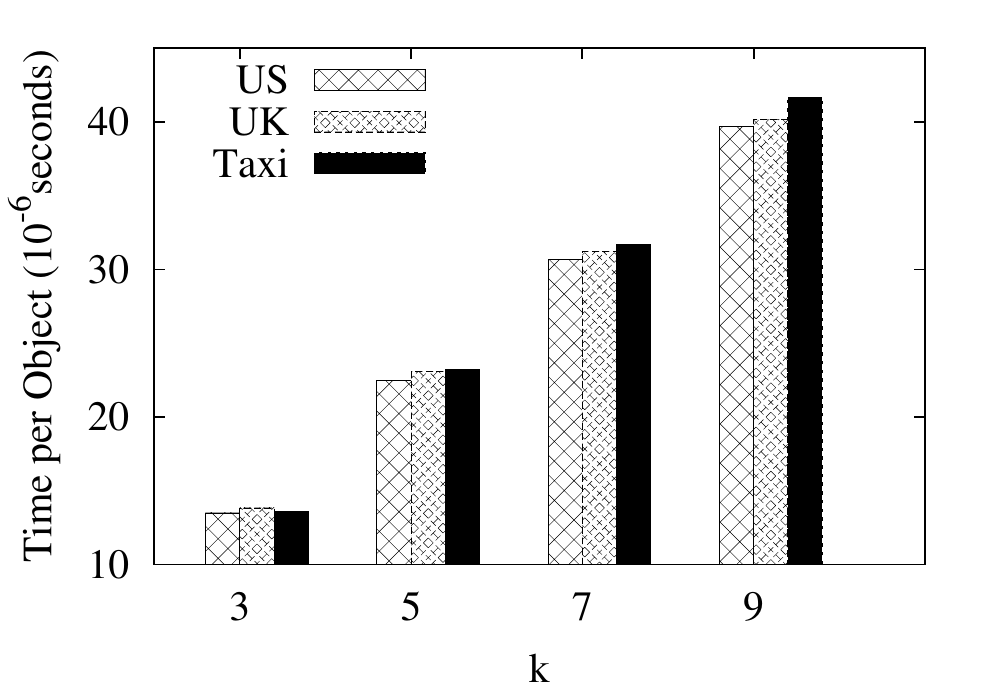}
}

\caption{Top-$k$ bursty regions detection.}
\label{fig:topkwindow}
\vspace{-4ex}\end{figure}

\subsection{Finding Top-k Bursty Regions}
We next evaluate the performance of the extensions of our three algorithms for continuously detecting top-$k$ bursty regions. We study the effect of $k$ and the size of sliding windows.

\vspace{1ex} \stitle{Runtime Performance.} This set of experiments aims to evaluate the efficiency of these algorithms \wrt the sliding window size. We adopt the same setting as in Section~\ref{sec:exactexp}. Figures~\ref{fig:topkwindow}(a)--(c) report the average runtime per object of  \textsf{kCCS},  \textsf{kGAPS}, and  \textsf{kMGAPS} for different sliding windows. We observe that as the sliding window gets larger, \textsf{kCCS} does not scale well and cannot process the top-$k$ queries efficiently. Meanwhile, \textsf{kGAPS} and \textsf{kMGAPS} can find top-$k$ bursty regions efficiently.

We also compare the na\"{i}ve solution for finding top-$k$ bursty regions with these algorithms. Recall from Section~\ref{sec:topk}, in the na\"{i}ve solution, we detect the top-$k$ bursty regions for each newly-arrived object. Clearly, the na\"{i}ve solution is prohibitively expensive. Hence, we only run it with a small sliding window on \textsf{US}, and its runtime per object is about 100X more than \textsf{kCCS}.

\vspace{1ex}\stitle{Effect of $k$.} Next, we study how the value of $k$ affect the runtime performance of the three extensions. We use the following 4 values for $k$: 3, 5, 7 and 9. The runtime performance is depicted in Figures~\ref{fig:topkwindow}(d)--(f). We observe the runtime per object of \textsf{kCCS} increases as $k$ increases. This is because we divide the top-$k$ bursty region detection problem into $k$ instances of bursty region detection problems. Each bursty region detection problem takes $O(n_c^2)$ time to find a bursty region, where $n_c$ is the number of spatial objects in the cells that we actually searched. In addition, we also observe that \textsf{kGAPS} and \textsf{kMGAPS} are less affected by $k$.

\subsection{Case Study}\label{appendix:casestudy}
In order to give a better view of our problem, we conduct a case study on the region monitored by our \textit{cell-}\textsc{cSpot} algorithm. Due to the space constraints, the detailed results are reported in Appendix~\ref{appendix:casestudy}.

\vspace{-1ex} \section{Conclusions}\label{sec:conclusion}
The work reported in this paper is motivated by new opportunities brought by the massive volumes of streaming geo-tagged data (i.e., spatial objects) generated by location-enabled mobile devices. Specifically, we have studied a new problem called the \brm problem to continuously detect the bursty region in a given area in real time. The \brm problem is important as it can underpin various applications such as disease outbreak detection.  We have proposed an exact solution and two approximate solutions for \brm. 
We have also extended these solutions to find top-$k$ bursty regions. Finally, our experiment study with real-world datasets has demonstrated the efficiency of our framework. As part of future work, we intend to explore the \brm problem in the context of road network.

\balance

\newpage
\bibliographystyle{abbrv}
\begin{small}
\bibliography{ref}
\end{small}
\appendix

\section{Proofs}\label{appendix:proofs}
\subsection{Proof for Theorem~\ref{the:reduce}}

\begin{proof}
Let $p$ be any point in the \bpm problem, and $r$ be the rectangular region of size $a\times b$ whose top-right corner is located at $p$. A spatial object $o$ is in $r$ iff the corresponding rectangle object $g$ can cover $p$. Since the corresponding $o$ and $g$ have the same creation time and weight, we can derive that $f(r, W_c) = f(p, W_c)$, $f(r, W_p) = f(p, W_p)$, and thus $r$ and $p$ have the same burst score. As a result, if the point $p_m$ has the maximum burst score in the \bpm problem, then $r_m$ also has the maximum burst score in the \brm problem.
\end{proof}

\subsection{Proof for Lemma~\ref{lemma:staticcorrect}}
\begin{proof} We have 
\begin{equation*}
\begin{split}
\mathcal{S}(p) & = \alpha \max(f(p, W_c) - f(p, W_p), 0) + (1-\alpha)f(p, W_c) \\
& \leq \alpha f(p, W_c) + (1-\alpha)f(p, W_c) = f(p, W_c) = U_{s}(c)
\end{split}
\end{equation*}
\end{proof}

\subsection{Proof for Lemma~\ref{lemma:hubcorrect}}

\begin{proof}
Let $\Delta \mathcal{S}(p)$, $\Delta f(p, W)$ be the increase of $\mathcal{S}(p)$ and $f(p, W)$ after $e$ happens, respectively. We discuss the following three cases.

\noindent\textbf{Case 1:  $e$ is new.}
For any point $p$ that is covered by $g$, its current score is increased by $\Delta f(p, W_c) = \frac{g.w}{|W_c|}$.
We have $\Delta \mathcal{S}(p) \leq \Delta f(p, W_c) = \frac{g.w}{|W_c|}$.

\noindent\textbf{Case 2: $e$ is grown.}
For any point $p$ that is covered by $g$, its current score is decreased, i.e., $\Delta f(p, W_c) = -\frac{g.w}{|W_c|}$, and its past score is increased, i.e., $\Delta f(p, W_p) = \frac{g.w}{|W_p|}$. Thus, we can easily get $\Delta \mathcal{S}(p) \leq 0$.

\noindent\textbf{Case 3: $e$ is expired.}
For any point $p$ covered by $g$, its current score is not affected, and its past score is decreased, i.e., $\Delta f(p, W_p) = -\frac{g.w}{|W_p|}$. Thus, we have $\Delta \mathcal{S}(p) \leq \alpha (-\Delta f(p, W_p)) = \alpha \frac{g.w}{|W_p|}$.

Since $\Delta \mathcal{S}(p) \leq \Delta U_d(c)$, we still have $\mathcal{S}(p) \leq U_d(c)$. 
\end{proof}

\subsection{Proof for Lemma~\ref{lemma:pointcorrect}}

\begin{proof}
We use $\Delta$ to denote the increase of the score. We consider the following three cases.

\noindent\textbf{Case 1: $e.l$ is new.} 
We have $\Delta \mathcal{S}(c.p) = \frac{g.w}{|W_c|}$ if and only if $g$ can cover $c.p$ and $f(c.p, W_c) - f(c.p, W_p) > 0$. In this case, $c.p$ still has the maximum burst score as $\Delta \mathcal{S}(p) \leq \frac{g.w}{|W_c|}$ for any $p$ in $g$ (Lemma~\ref{lemma:hubcorrect}). Otherwise, it is possible that there exists a point $p'$ in $g$ with a larger increase such that $p'$ has a larger burst score than $c.p$ after $g$ arrives.

\noindent\textbf{Case 2: $e.l$ is grown.} 
For any point $p$ in $g$, the increase $\Delta \mathcal{S}(p) < 0$. If $g$ does not cover $c.p$, $c.p$'s burst score does not change and it still has the maximum burst score. Otherwise, $c.p$'s burst score is decreased and could be exceeded by a point outside $g$.

\noindent\textbf{Case 3:$e.l$ is expired.} 
As shown in the proof for Lemma~\ref{lemma:hubcorrect}, $\Delta \mathcal{S}(p) \leq \alpha \frac{g.w}{|W_c|}$ for any $p$ in $g$. We have $\Delta \mathcal{S}(c.p) = \alpha \frac{g.w}{|W_c|}$ if and only if $g$ can cover $c.p$ and $f(c.p, W_c) - f(c.p, W_p) > 0$. In this case, $c.p$ still has the maximum burst score.
Otherwise, similar to Case 1, it is possible that there exists a point $p'$ in $g$ with a larger increase of burst score.

Putting these together, the lemma is proved.
\end{proof}

\subsection{Proof for Lemma~\ref{lemma:scorefunction2}}
\begin{proof}
According to the definition of the burst score, we have
\begin{equation*}
\begin{split}
&\mathcal{S}(r_2)
 =  \alpha \max(f(r_2, W_c) - f(r_2, W_p), 0) + (1-\alpha) f(r_2, W_c) \\
 \geq & (1-\alpha) f(r_2, W_c) \geq (1-\alpha) f(r_1, W_c) \geq (1-\alpha) \mathcal{S}(r_1)
\end{split}
\end{equation*}
\end{proof}

\subsection{Proof for Lemma~\ref{lemma:scorefunction3}}
\begin{proof}
Since $r_1$ and $r_2$ are non-overlapping, according to the definition of burst score, we have
$$f(r_1, W_c) + f(r_2, W_c) = f(r_1\cup r_2, W_c)$$
$$f(r_1, W_p) + f(r_2, W_p) = f(r_1\cup r_2, W_p)$$
Then we can easily get
\begin{equation*}
\begin{split}
&\max(f(r_1\cup r_2, W_c) - f(r_1\cup r_2, W_p), 0) \\
\leq & \max(f(r_1, W_c)-f(r_1, W_p), 0) + \\
&\max(f(r_2, W_c)-f(r_2, W_p), 0)
\end{split}
\end{equation*}
Thus, we have $\mathcal{S}(r_1\cup r_2) \leq \mathcal{S}(r_1) + \mathcal{S}(r_2)$.
\end{proof}

\subsection{Proof for Theorem~\ref{theorem:gridbound}}
\begin{figure}[t]
\minipage{0.5\linewidth}
\centering
\vspace{3ex}
\includegraphics[width=0.8\linewidth]{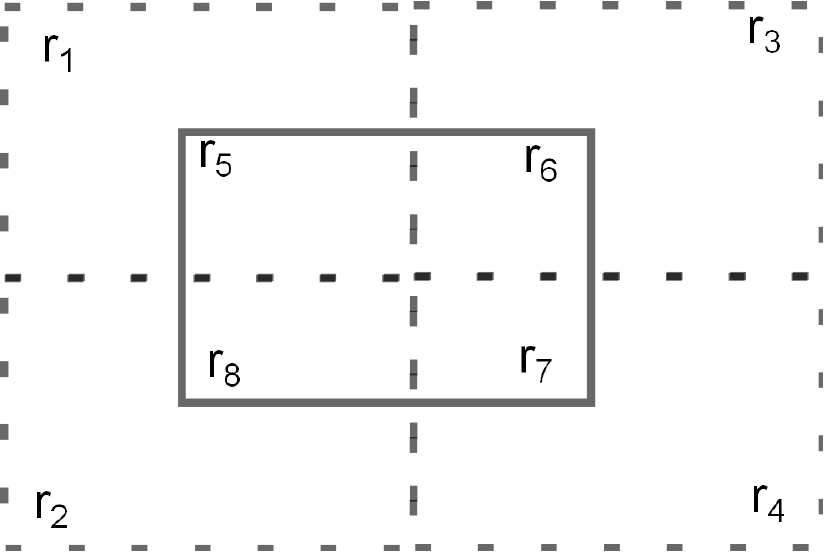}
\vspace{-1ex}
\caption{Proof for Theorem~\ref{theorem:gridbound}}\label{fig:gridapproxbound}
\endminipage
\minipage{0.5\linewidth}
\centering
\vspace{3ex}
\includegraphics[width=0.8\linewidth]{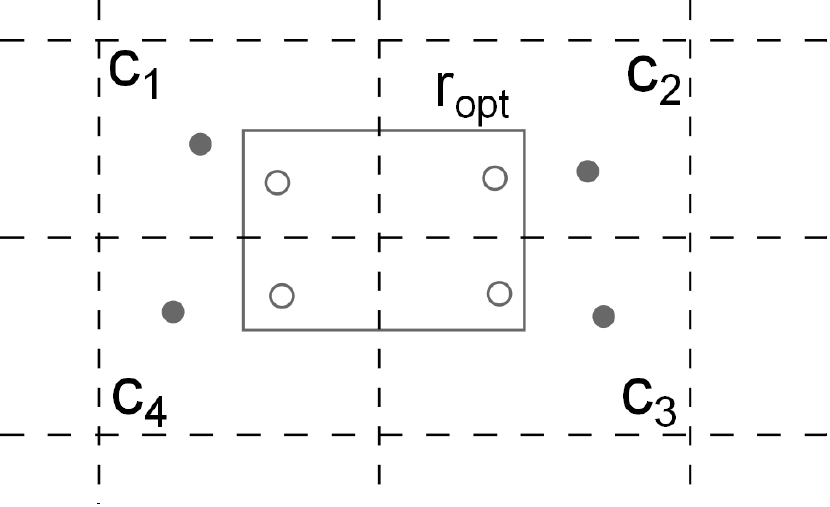}

\caption{A tight example.}\label{fig:tightbound}
\endminipage
\vspace{-2ex}
\end{figure}

\begin{proof}
Since the sizes of $r_{opt}$ and any cell are both $a\times b$, then $r_{opt}$ either overlaps with a cell or intersects with four cells. We consider the following two cases.

\noindent\textbf{Case 1: $r_{opt}$ overlaps one cell.} Since we return the candidate with maximum burst score, we will return the bursty region to users. The approximate ratio is 1.

\noindent\textbf{Case 2: $r_{opt}$ intersects with 4 cells.} Consider the example shown in Figure~\ref{fig:gridapproxbound}. Let the solid line rectangle be $r_{opt}$. The four dashed line rectangles are four cells which intersect with $r_{opt}$. According to Lemma~\ref{lemma:scorefunction2}, we have $(1-\alpha) \mathcal{S}({r_{opt}}) \leq \mathcal{S}(c_1\cup \dots \cup c_4)$. According to Lemma~\ref{lemma:scorefunction3}, we can derive that $\mathcal{S}(c_1\cup \dots \cup c_4) \leq \sum_{i\in[1,4]}\mathcal{S}({c_i})$. Since we report the cell with the maximum burst score, i.e., $\mathcal{S}(r) \geq \mathcal{S}(c_i)$ for any $i\in[1,4]$. Thus, we have $\frac{1-\alpha}{4}\mathcal{S}({r_{opt}}) \leq \mathcal{S}({r})$.
\end{proof}

\subsection{Proof for Lemma~\ref{lemma:tight}}
\begin{proof}
We show the approximation ratio is tight by giving an example. Consider an instance in Figure~\ref{fig:tightbound}, where $c_1, c_2, c_3$ and $c_4$ are cells in the grid, and the solid-line rectangle $r_{opt}$ is the bursty region with the maximum burst score. The white nodes are the spatial objects in window $W_c$ and the black nodes are in $W_p$. We assume that $\frac{o.w}{|W_c|} = \frac{o.w}{|W_p|} = 1$ for each object $o$. The burst score for the region $r_{opt}$ is $\mathcal{S}({r_{opt}}) = \alpha \max(4-0, 0)+ (1-\alpha) 4 = 4$.
The burst score of cell $c_i$ is $\mathcal{S}({c_i}) = \alpha \max(1-1, 0) + (1-\alpha) = 1-\alpha$, for any $i\in [1, 4]$. Thus, the approximation ratio is tight.
\end{proof}

\subsection{Proof for Theorem~\ref{theorem:multigrid}}
\begin{proof}
Since the \textsc{mGap-surge} returns the best result of found by Algorithm~\ref{alg:gridupdate}, its approximation ratio is $\frac{1-\alpha}{4}$.
\end{proof}

\section{Pseudocode for MGAP-surge}\label{appendix:mgap-surge}
The pseudocode for the \textsc{mGap-surge} Algorithm is presented in Algorithm~\ref{alg:4grid}.
\begin{algorithm}[h]
\caption{\textsc{mGap-surge} Algorithm}
\label{alg:4grid}
\begin{small}
\SetArgSty{textnormal}
\KwIn{Spatial Object $o$}
\KwOut{A region of size $a\times b$ }
$r_1 \gets$ \FuncSty{\textsc{gap-surge}($Grid1, o, H_1$)}\;
$r_2 \gets$ \FuncSty{\textsc{gap-surge}($Grid2, o, H_2$)}\;
$r_3 \gets$ \FuncSty{\textsc{gap-surge}($Grid3, o, H_3$)}\;
$r_4 \gets$ \FuncSty{\textsc{gap-surge}($Grid4, o, H_4$)}\;
\Return{The region among $r_1 \dots, r_4$ with highest burst score.}
\end{small}
\end{algorithm}

\section{Pseudocode of Top-$k$ Bursty Regions Detection Algorithms}\label{appendix:algorithm}
\begin{algorithm}[h]
\caption{\textsc{Gap-Ksurge} Algorithm}
\label{alg:apptopk}
\begin{small}
\SetArgSty{textnormal}
\KwIn{Spatial Object $o$, Heap $H$, integer $k$, $Grid$}
\KwOut{A region of size $a\times b$ }
$c_{i, j} \leftarrow $ the cell $o$ lies in\;
\If{$c_{i,j}$ not in $H$}{
    add $c_{i,j}$ to $H$\;
}
\If{$o.t_c \in W_c$}{
    $r_{can}.S_c += \frac{o.\rho}{|W_c|}$\;
}
\ElseIf{$o.t_c \in W_p$}{
    $r_{can}.S_c -= \frac{o.\rho}{|W_c|}, r_{can}.S_p += \frac{o.\rho}{|W_p|}$\;
}
\Else{
    $r_{can}.S_p -= \frac{o.\rho}{|W_p|}$\;
}
$r_{can}.S = \max(r_{can}.S_c - r_{can}.S_p, 0) + r_{can}.S_c$\;
update $H$\;
\Return{Top-$k$ cells in $H$}
\end{small}
\end{algorithm}

\begin{algorithm}[h]
\caption{\textsc{mGap-KSurge} Algorithm}
\label{alg:4gridtop4}
\begin{small}
\SetArgSty{textnormal}
\KwIn{Spatial Object $o$}
\KwOut{A region of size $a\times b$ }
$l_1 \gets$ \FuncSty{\textsc{gap-surge}($Grid1, o, H_1, 4k$)}\;
$l_2 \gets$ \FuncSty{\textsc{gap-surge}($Grid2, o, H_2, 4k$)}\;
$l_3 \gets$ \FuncSty{\textsc{gap-surge}($Grid3, o, H_3, 4k$)}\;
$l_4 \gets$ \FuncSty{\textsc{gap-surge}($Grid4, o, H_4, 4k$)}\;
$r[1, k] \gets $ top-$k$ non-overlapping cells from $l_1 \cup l_2 \cup l_3 \cup l_4$\;
\Return{$r[1, k]$}

\end{small}
\end{algorithm}

\section{Additional Experiments}\label{appendix:additional}

\subsection{Details of the evaluated algorithms}\label{appendix:evaluatedmethods}
We evaluate the performances of the three proposed algorithms, namely the exact method \textit{Cell}-\textsc{cSpot} (denoted by \textsf{CCS}), the grid-based approximation algorithm \textsc{gap-surge} (denoted by \textsf{GAPS}), and the multi-grid-based technique \textsc{mGap-surge} (denoted by \textsf{MGAPS}).  We denote the top-$k$ extensions of these algorithms as  \textsf{kCCS},  \textsf{kGAPS}, and  \textsf{kMGAPS}, respectively. To evaluate the usefulness of our proposed method of upper bound estimation, we compare \textsf{CCS} with an approach that only utilizes the static upper bound. We denote this baseline method by \textsf{B-CCS}. We also compare \textsf{CCS} with a baseline approach that does not use any upper bound estimation technique, denoted by \textsf{Base}. Specifically, in \textsf{Base} we divide the space into cells, and we search all the cells that overlap with the rectangle object when an event happens. To the best of our knowledge, there is no existing technique that address the \brm problem. Hence we are confined to compare our proposed algorithms with \textsf{aG2} \cite{amagata2016monitoring}, which is designed for continuously monitoring the
MaxRS problem. Obviously, we cannot directly apply it to solve the \brm problem. In our experiments, we use a modified version of \textsf{aG2}. Specifically, the modified algorithm inherits the grid index structure and the branch-and-bound strategy from the original algorithm. The main difference between the modified and the original algorithms is how we search a rectangle object given a snapshot of the stream. In the original algorithm, they invoke the sweep-line algorithm \cite{nandy1995unified} to search a rectangle object to find a region with maximum sum score, while in the modified algorithm, we use our proposed \textsc{sl-cSpot} algorithm instead. 

\subsection{Approximate Ratio}\label{appendix:ratio}
\vspace{1ex}\stitle{Approximate Ratio.} In this set of experiments, we vary the sliding window to assess the approximate ratio of the burst scores of region detected by \textsf{GAPS} and \textsf{MGAPS}. The detailed results are reported in Table~\ref{tab:approxratio_window}. Though the theoretical approximate ratio is $\frac{1-\alpha}{4}$, in practice it is much better, especially for \textsf{MGAPS}. We observe that for \textsf{UK}, the burst score of the region detected by \textsf{GAPS} is about 70\%--90\% of the burst score of the optimal region.  The region detected by \textsf{MGAPS} is about 85\%--95\% of the burst score of the optimal region. Since \textsf{GAPS} and \textsf{MGAPS} are much more efficient than \textsf{CCS} (about three orders of magnitude faster), they are good alternatives to \textsf{CCS} when a slight imprecision
is acceptable.

\begin{table}[t]
\caption{Approximate ratio vs. the size of window.} \label{tab:approxratio_window}
\vspace{-3ex}
\begin{center}
\begin{scriptsize}
\centering
\begin{tabular}[h]{| c | c | c | c | c | c | c | }
\hline
\multirow{3}{*}{Taxi} & Window (mins) & 1  &  5  & 10  & 20  & 30  \\ \cline{2-7}
& \textsf{GAPS} & 76.34\% & 73.90\% & 75.12\% & 75.70\% & 76.35\% \\ \cline{2-7}
& \textsf{MGAPS} & 85.98\% & 85.14\% & 87.35\% & 88.34\% & 87.85\% \\ \hline \hline
\multirow{3}{*}{UK} & Window (hours) & 0.5  & 1 & 2  & 12  & 24  \\ \cline{2-7}
& \textsf{GAPS} & 90.22\% & 91.56\% & 91.98\% & 89.82\%  & 92.44\% \\ \cline{2-7}
& \textsf{MGAPS} & 93.13\% & 94.34\% & 93.76\% & 90.50\% & 92.82\% \\ \hline \hline
\multirow{3}{*}{US} & Window (hours) & 0.5  & 1  & 2  & 12  & 24  \\ \cline{2-7}
& \textsf{GAPS} & 84.23\% & 80.67\% & 89.70\% &  91.77\% &  80.10\% \\ \cline{2-7}
& \textsf{MGAPS} & 88.61\% & 88.07\%b & 91.44\%  & 91.77\% & 84.34\% \\ \hline
\end{tabular}
\end{scriptsize}
\vspace{-2ex}
\end{center}
\vspace{-2ex}
\end{table}

\subsection{Case Study}\label{appendix:casestudy}
To evaluate the result quality of our \textit{cell-}\textsc{cSpot} algorithm, we conduct a case study on the region monitored by the algorithm. We run the \textit{cell-}\textsc{cSpot } algorithm on the tweets posted in United States from 2012 April to 2012 October. Note that since the algorithm continuously reports the location of bursty regions, we only present two examples of the detected bursty region and explain the connection between the region and real life events.

\begin{figure}[h]
\minipage{0.9\linewidth}
  \centering
  \includegraphics[width=\linewidth]{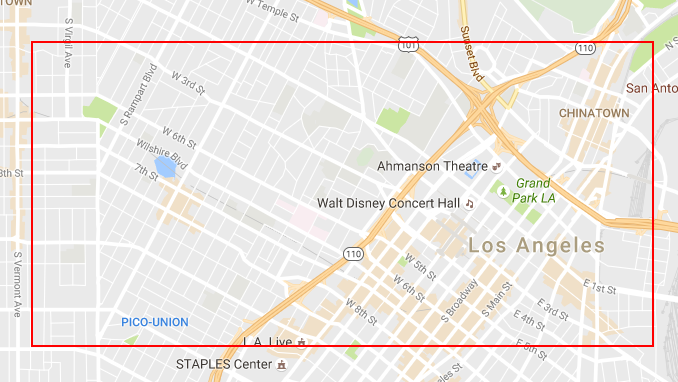}
 \caption{Bursty Region about ``concert''.}\label{fig:case1}
 \endminipage
 \hspace{1ex}
 \minipage{0.9\linewidth}
   \centering
   \includegraphics[width=\linewidth]{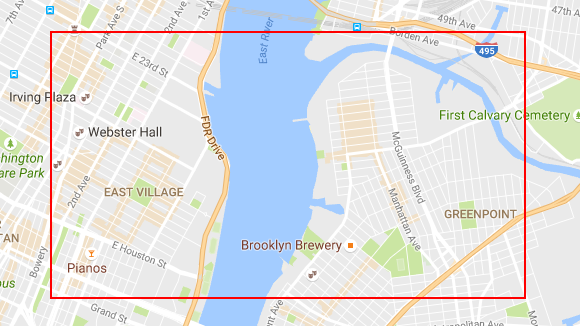}
  \caption{Bursty Region about ``parade''.}\label{fig:case2}
 \endminipage
\vspace{-2ex} \end{figure}

\begin{example}
  In the first example, we present detecting bursty regions about ``concert''. Specifically, we only consider tweets containing keyword ``concert'' and continuously report the detected bursty region. On July 8, 2012, our algorithm
detected a region as shown in Figure~\ref{fig:case1}. The frequent keywords in this region during this time are ``Walt'' and ``Concert''.
By checking the events that happened in July 2012, we find that there was a concert performed by Ketherine Eason with Inner City Youth Orchestra of Los Angeles in Walt Disney Concert Hall in the detected region. 
\end{example}

\begin{example}
In the second example, we present detecting bursty regions about ``parade''. On May 19, 2012, our algorithm detected a region as shown in Figure~\ref{fig:case2}. The frequent keywords in this region are ``annual'', ``dance'', and ``parade''. By checking the events that happened in May 2012, we noticed that the dance parade is an annual parade and festival in New York. Specifically, in 2012 the parade took over Broadway Street on May 19th.
\end{example}

\end{document}